\documentclass[twocolumn]{aastex631}
\usepackage{footnote, natbib}

\usepackage{color}

\definecolor{titlecol1}{rgb}{0.039,0.361,0.569} 
\definecolor{titlecol2}{rgb}{0.7,0.0,0.05} 

\begin{document}

\title{Harnessing the Hubble Space Telescope Archives: A Catalogue of 21,926 Interacting Galaxies}

\correspondingauthor{David O'Ryan}
\email{d.oryan@lancaster.ac.uk}

\author[0000-0003-1217-4617]{David O'Ryan}
\affiliation{Department of Physics, Lancaster University, Bailrigg, Lancaster, LA1 4YB, UK}
\affiliation{European Space Agency (ESA), European Space Astronomy Centre (ESAC), Camino Bajo del Castillo s/n, 28692, Villaneuva de la Ca\~nada, Madrid}

\author[0000-0002-8555-3012]{Bruno Mer\'in}
\affiliation{European Space Agency (ESA), European Space Astronomy Centre (ESAC), Camino Bajo del Castillo s/n, 28692, Villaneuva de la Ca\~nada, Madrid}

\author[0000-0001-5882-3323]{Brooke D. Simmons}
\affiliation{Department of Physics, Lancaster University, Bailrigg, Lancaster, LA1 4YB, UK}

\author{Ant\'onia Vojtekov\'a}
\affiliation{European Space Agency (ESA), European Space Astronomy Centre (ESAC), Camino Bajo del Castillo s/n, 28692, Villaneuva de la Ca\~nada, Madrid}

\author{Anna Anku}
\affiliation{European Space Agency (ESA), European Space Astronomy Centre (ESAC), Camino Bajo del Castillo s/n, 28692, Villaneuva de la Ca\~nada, Madrid}

\author[0000-0002-6408-4181]{Mike Walmsley}
\affiliation{Jodrell Bank Centre for Astrophysics, Department of Physics and Astronomy, University of Manchester, Oxford Road, Manchester, M13 9PL, UK}

\author[0000-0002-3887-6433]{Izzy L. Garland}
\affiliation{Department of Physics, Lancaster University, Bailrigg, Lancaster, LA1 4YB, UK}

\author[0000-0002-6851-9613]{Tobias G\'eron}
\affiliation{Oxford Astrophysics, Department of Physics, University of Oxford, Denys Wilkinson Building, Keble Road, Oxford, OX1 3RH, UK}

\author{William Keel}
\affiliation{Department of Physics and Astronomy, University of Alabama, Box 870324, Tuscaloosa, AL 35487}

\author[0000-0001-8010-8879]{Sandor Kruk}
\affiliation{Max-Planck-Institut f\"ur Extraterrestrische Physik (MPE), Giessenbachstrasse 1, D-85748 Garching bei M\"unchen, Germany}

\author[0000-0001-5578-359X]{Chris J. Lintott}
\affiliation{Oxford Astrophysics, Department of Physics, University of Oxford, Denys Wilkinson Building, Keble Road, Oxford, OX1 3RH, UK}

\author[0000-0002-6016-300X]{Kameswara Bharadwaj Mantha}
\affiliation{School of Physics and Astronomy, University of Minnesota, 116 Church Street SE, Minneapolis, MN 55455, USA}

\author[0000-0003-0846-9578]{Karen L. Masters}
\affiliation{Departments of Physics and Astronomy, Haverford College, 370 Lancaster Avenue, Haverford, Pennsylvania 19041, USA}

\author{Jan Reerink}
\affiliation{European Space Agency (ESA), European Space Astronomy Centre (ESAC), Camino Bajo del Castillo s/n, 28692, Villaneuva de la Ca\~nada, Madrid}

\author[0000-0001-6417-7196]{Rebecca J. Smethurst}
\affiliation{Oxford Astrophysics, Department of Physics, University of Oxford, Denys Wilkinson Building, Keble Road, Oxford, OX1 3RH, UK}

\author[0000-0003-1014-5839]{Matthew R. Thorne}
\affiliation{Department of Physics, Lancaster University, Bailrigg, Lancaster, LA1 4YB, UK}

\begin{abstract}
  Mergers play a complex role in galaxy formation and evolution. Continuing to improve our understanding of these systems require ever larger samples, which can be difficult (even impossible) to select from individual surveys. We use the new platform ESA Datalabs to assemble a catalogue of interacting galaxies from the \emph{Hubble Space Telescope} science archives; this catalogue is larger than previously published catalogues by nearly an order of magnitude. In particular, we apply the \texttt{Zoobot} convolutional neural network directly to the entire public archive of \emph{HST} $F814W$ images and make probabilistic interaction predictions for 126 million sources from the \emph{Hubble} Source Catalogue. We employ a combination of automated visual representation and visual analysis to identify a clean sample of 21,926 interacting galaxy systems, mostly with $z < 1$. 65\% of these systems have no previous references in either the NASA Extragalactic Database or Simbad. In the process of removing contamination, we also discover many other objects of interest, such as gravitational lenses, edge-on protoplanetary disks, and `backlit' overlapping galaxies. We briefly investigate the basic properties of this sample, and we make our catalogue publicly available for use by the community. In addition to providing a new catalogue of scientifically interesting objects imaged by \emph{HST}, this work also demonstrates the power of the ESA Datalabs tool to facilitate substantial archival analysis without placing a high computational or storage burden on the end user.
\end{abstract}



\section{INTRODUCTION}\label{intro}
\noindent Interacting and merging galaxies are important to our current theory of $\Lambda$CDM cosmology, in which structure typically assembles hierarchically \citep{abadi_03, springel05, delucia_07, guo_08}. Galaxy interaction leads to highly disturbed morphologies \citep{toomre72, hernandez05, wallin15}, intense starbursts \citep{mihos96, springel00, Saitoh09, moreno21} and, potentially, quenching of some systems \citep{hopkins13, smethurst18, hani20, das22}. In general, galaxies undergoing interaction are observed to have higher star formation rates than those that exist in the field \citep{ellison_08, scudder12, pearson19b}. Interaction also has a direct impact on the gas angular momentum within each galaxy, causing it to decrease. This, potentially, leads to funnelling of gas into into their nuclear regions and igniting activity. This could be a connection with active galactic nuclei \citep{ellison_08, li08, ellison_11, comerford15}. However, such a connection remains debated \citep{alonso07, mckernan10, marian20}. Thus, understanding galaxy interaction is crucial to testing theories of galaxy evolution itself.

Interacting galaxies have long been explored with different samples of galaxies. Examples include constraining merger rates as a function of redshift \citep{lotz08}, inferring the contribution of minor mergers to the cosmic star formation budget \citep{kaviraj14a, kaviraj14b}, and examining interactions as a function of their local environments, internal properties and AGN activity \citep{darg10}. These studies \citep[and many others; for further examples, see][]{barton00, alonso04, ellison13, holincheck16, silva21} illustrate the complex parameter space involved in understanding the role of interaction in galaxy evolution. Thus, to effectively study interacting galaxies, we need observed datasets of such a size that they can sample a wide range of various parameters of interest.

The first large-scale catalogues of interacting galaxies are from the mid 20th century \citep[][hereafter VV]{arp_cat, vv_cat, vv_catb}. These catalogues primarily used visual inspection to identify mergers \citep[e.g.,][]{demello97, nair10} and generally found from hundreds to thousands of systems. The largest set of interacting galaxies identified by a single expert classifier contains 2,565 relatively nearby systems \citep{arp87}. Citizen science techniques can extend this number, as was presented by \citet{darg_cat} who used them to find a catalogue of 3,003 interacting galaxies.

The inclusion of automated classification shows promise to continue this expansion. The use of machine learning in classifying galaxy morphology is well established \citep{ardizzone96, elaziz17, barchi20, ghosh20, cheng21}. The workhorse algorithm is the convolutional neural network \citep[CNN; for an introduction, see][]{oshea15}, most often used in image recognition and feature extraction. CNNs can be used for general classification (e.g. early- versus late-type galaxies) or to extract specific morphological features of galaxies, such as bars, spiral arms, etc; many works have demonstrated their effectiveness at this \citep[e.g.][]{ackermann_18, jacobs_19, bickley_21, buck_21, walmsley22}. \citet{Pearson_22} demonstrated the power of CNNs for finding interacting and merging galaxies specifically, finding 2,109 in 5.4 deg$^2$ of Hyper Suprime-Cam imagery - a large sample for the small area covered.

However, issues with using CNNs in classifying interacting galaxies have been found on numerous occasions. The primary concern, is that - without due care - classifying interacting galaxies by morphology alone can be highly contaminated. For example, CNNs often confuse chance alignments of galaxy pairs on the sky for interacting systems. This leads to many predicted interacting systems being thrown away after visual inspection (in some cases up to 60\%; \citet{bottrell_19, Pearson_22}). 


In this work, we aim to use machine learning to create a large, high-confidence catalogue of interacting systems, drawn entirely from existing astronomical imagery. We search through the European Space Agency's \emph{Hubble Space Telescope} Science Archive\footnote{See \url{http://hst.esac.esa.int/ehst/}} using a CNN to predict whether an image contains an interacting system, from among the 126 million extended objects in the \emph{Hubble} Source Catalogue \citep[HSC;][]{whitmore_16}. The feature extraction we implement is focused on finding tidal features or morphological disturbance caused by the interaction. The tidal features prioritised include tidal tails, tidal bridges or tidal debris. As stated previously, this runs the risk of introducing high levels of contamination by close pairs. We thus implement further automated and manual methods, which significantly reduce this. The systems we find are often in the background of previous deep surveys (such as the Cosmic Evolution Survey, COSMOS, \citealt{scoville07}; the Great Observatories Origins Deep Survey, GOODS, \citealt{giavalisco04}; and the Pancromatic Hubble Andromeda Treasury Survey, PHAT, \citealt{dalcanton12}), where spectroscopic coverage varies. Therefore, while our final catalogue reduces contamination to $\sim 3\%$, definitively removing all contamination by close pairs remains a challenge following this work.

This paper is laid out as follows: Section \ref{data} describes the HSC and all the criteria we applied to create the images we predict over. This Section also introduces ESA Datalabs\footnote{\url{https://datalabs.esa.int/}}; a new platform which allows the user to directly access the \emph{Hubble} Science Archive. Section \ref{CNNs} gives an in depth description of the \texttt{Zoobot} CNN we utilise for our predictions, and how it differs from a commonly used CNN. Section \ref{training} explains the process of creating the training set for our CNN to find interacting galaxies, with Section \ref{diagnostics} showing how well it performed and providing the diagnostics of the CNN. We also use this Section to investigate the contamination in our catalogue. Section \ref{results} describes our results and discusses the final catalogue as well as define interesting systems or objects that we have found. We also explore some basic properties of the catalogue here. Finally, Section \ref{conclusion} summarises our results and conclusions.

Where necessary, we use a Flat $\Lambda$CDM cosmology with $H_0$ = $70$\,km/s/Mpc and $\Omega_M = 0.3$. Hereafter in this paper, when referring to an interacting galaxy we are referring to a galaxy which has undergone one or multiple flybys by a secondary galaxy and caused tidal disturbance. A merging galaxy is the final state of these flybys, where two or more systems have coalesced to form a highly morphologically irregular system.


\section{DATA}\label{data}
\subsection{The \emph{Hubble} Archives \& ESA Datalabs}
\noindent The observational data is directly from the \emph{Hubble} Science Archive and is accessed from the new ESA Datalabs platform. The repository contains approximately 100TB of data from the \emph{Hubble Space Telescope} (\emph{HST}). This repository spans all \emph{HST} instruments and filters. ESA Datalabs provides a direct interface between users and the data. On this platform, every observations' FITS file can be accessed. To streamline our pipeline, we applied criteria to the observations as not all filters have the same number of observations, some instruments are not as sensitive to the low surface brightness regime as others or the field of view of certain instruments would not be ideal for measuring galaxy morphology. Finally, we do not conduct source extraction from each FITS file ourselves but use the \emph{Hubble} Source Catalogue \citep[][hereafter HSC]{whitmore_16} to define the centre of each source cutout.

The criteria we apply are: the observational data must be from the Advanced Camera for Surveys (ACS), it must be final product data of \emph{HST} (i.e. within a .drc file, where the data has been drizzle \citep{drizzlepac} combined and had charge-transfer-efficiency corrections applied), observed within the $F814W$ filter and must be flagged as an extended source in the HSC. This offloads sky subtraction, cosmic ray rejection and charge efficiency calculations to the original \emph{HST} pipeline and removes costly steps from our cutout creation process. We utilise all final product data of the $F814W$ filter from \emph{HST} as this was the filter which contained the most FITS files, and therefore observations. The $F814W$ filter contained 9,527 final product FITS files which could be used for source extraction, whereas the closest second (the $F606W$ filter) contained $\approx$6000. By using the filter with the most files, we are confident that we cover a majority of the HSC. Applying this criteria gives 126 million sources to predict over.

We must create 126 million source cutouts from 9,507 different FITS files. Creating a dataset of cutouts at this magnitude in conventional methods (such as \texttt{AstroQuery} or Table Access Protocol (TAP) services) would be impractical due to making many network calls and long FITS file download times. Instead, we use the ESA Datalabs platform, which is due to be released in Q3 of 2023. This platform has been developed to allow us to `mount' the \emph{Hubble} Science Archive onto it. In practice, providing access to the entire \emph{Hubble} Science Archives as local files for the user to manipulate while on the platform. This bypasses network calls to servers to download our required FITS files, a process which could have taken minutes per download. Having direct access to the files, and quickly matching source coordinates to FITS files (described in Section \ref{benchmarks}) allows us to open a FITS file and create all source cutouts from it without having to close or reopen it. Therefore, we were able to create on the order of 10k cutouts in the same order of time taken to download a single file.

The source cutouts were created as $F814W$ gray scaled 150x150 (7.5$^{\prime \prime}$x7.5$^{\prime \prime}$) pixel images using the HSC source coordinates as the centre. The image size was set and standardized to streamline the pipeline. The majority of cutouts are centered on the source but, in a minority, misalignment between source and image centre occurs. This is a result of the drizzling process, with incorrect alignment sometimes being significant. However, the target source was always present in the cutout and we, therefore, did not attempt to rectify this. A ZScaleInterval with a hard set contrast of 0.05 and a LinearStretch following the default parameters in the \texttt{Astropy} \citep{astropy:2013, astropy:2018} package. These were binned to 300x300 pixels (pixel resolution is 3.25$^{\prime \prime}$x3.25$^{\prime \prime}$) with a linear interpolation from the \texttt{CV2} python package. The images were created at 150$\times$150 to minimise storage required on the early version of ESA Datalabs being used. Creating the images at half the size allowed us to scale up to 300$\times$300 pixels without any effects of the interpolation.

\subsection{The Shapely Python Package}\label{benchmarks}
\noindent A large computational expense in our pipeline was matching FITS files to sources. Conventionally, the \texttt{Astropy} \textsc{contains} function would be used to match source coordinates to the FITS file WCS. We instead use the \texttt{Shapely}\footnote{\texttt{Shapely} docs: \url{https://shapely.readthedocs.io/en/stable/manual.html}} Python package.  \texttt{Shapely} is a geometry orientated package primarily focused on geospatial data. We found converting the FITS image footprints into \texttt{Shapely} Polygons and the source coordinates to \texttt{Shapely} Points and then checking if they overlapped had significant speed up. Per iteration, Astropy's \textsc{contained\_by} function matches a source to a FITS file on the order of 500ms. Using \texttt{Shapely's} \textsc{contains} function, the same process is on the order of 6$\mu$s.


\section{UTILISING A CONVOLUTIONAL NEURAL NETWORK}\label{CNNs}
\noindent We must choose a CNN which would best suit our needs to classify them into interacting galaxies or not. We select the newly developed CNN \texttt{Zoobot}\footnote{\texttt{Zoobot} DOI: 10.5281/zenodo.6483176} \citep{walmsley22}. \texttt{Zoobot} is a CNN specifically trained to classify galaxies based on morphology into many different types (spiral, disk, elliptical, barred, non-barred, etc). We retrain it to only classify galaxies into interacting or non-interacting. Instead of training \texttt{Zoobot} from scratch and creating a new model, we use transfer learning to finetune existing \texttt{Zoobot} models to classify our data for our particular question. This allows us to retain information from \texttt{Zoobot's} previous training. More importantly, it requires a significantly smaller training set to achieve high accuracy.

\subsection{Zoobot}
\noindent The version of \texttt{Zoobot} we use is a deep CNN which was trained on Galaxy Zoo volunteer classifications over three different Galaxy Zoo: DECaLS (GZD)\citep[Dark Energy Camera Legacy Survey, described in][]{desi:paper} campaigns. These were GZD-1, GZD-2 and GZD-5 - each number corresponding to the DECaLS data release. For training \texttt{Zoobot}, DECaLS imaging was selected using the NASA-Sloan Atlas (NSA), which was itself constructed with SDSS Data Release 8 (DR8) images. This also introduced implicit cuts to the training data, as SDSS can not get to the depths of DECaLS. This introduces implicit magnitude and redshift cuts on the training data. Specifically, SDSS DR8 and the NSA cover galaxies brighter than m$_r > 17.77$ and closer than z $< 0.15$. In Section \ref{trans_learn} we describe using transfer learning to use Zoobot effectively outside of this magnitude and redshift range.

\citet{walmsley22} use the 249,581 volunteer classifications from GZD-5 campaign to train \texttt{Zoobot} to answer all 34 questions \citep[example shown in Figure 4 of][]{walmsley22} in the remaining campaigns. GZD-5 was used as it had a slightly different volunteer decision tree, having an expanded question on potential different galaxy merger stages. Each galaxy image had been shown to volunteers as a 3-color (g,r,z) of 424$\times$424 cutout. Each images pixel scale was an interpolation between the measured Petrosian 50\%- and 90\%-light radius. The measured full Petrosian radius had to be at least 3$^{\prime\prime}$ to be shown to the volunteers. When inputting into \texttt{Zoobot}, these cutouts were scaled and grayscaled to 300$\times$300$\times$1 images, averaging over the 3-color channels to remove colour information and avoid biasing the morphology predictions. \texttt{Zoobot} utilised the Adam \citep{kingma15} optimizer to train.

By training \texttt{Zoobot} in this way, combining the approach of answering many questions at once with Bayesian representation learning, it learns a generalisable summary of many types of galaxies. These generalised summarys are lower-dimensional descriptions of galaxy types and are referred to as representations. These representations change depending on the galaxy type, morphology or environment in an image and lead to similar images being closer together in a representation space than dissimilar ones. This representation approach on a very broad classification problem is found to increase accuracy and generality of \texttt{Zoobot}, giving it an edge over conventional CNNs. A more detailed breakdown of this approach, as well as further details about \texttt{Zoobot}s' architecture, can also be found in \citet{walmsley22}.

\texttt{Zoobot} was trained to give a prediction score to an image of a galaxy based on the question it is answering. The type of prediction score is set by the users choice of the model final layer in \texttt{Zoobot}. We elect to use a \textsc{SoftMax} output, which returns an output score as a float between 0 and 1. This prediction score is not a probability score, although it may seem analogous. A well behaved prediction score will map to probability, though not necessarily linearly. The mapping between prediction score and probability is not considered in this work, and we use the prediction score as an indicator of \texttt{Zoobot}'s confidence a source is an interacting galaxy. 

We are only interested in the `Is the galaxy merging or disturbed?' question from the Galaxy Zoo: DECaLS workflow, where the answer can be `merging', `major disturbance', `minor disturbance' or `None', and only want our version of \texttt{Zoobot} to return the answer to this. Our version of \texttt{Zoobot} is also not trained to predict over \emph{HST} data which differs from DECaLS data (different resolutions, filter bandwidths, etc). If we were to use our version of \texttt{Zoobot} as downloaded we would likely lose accuracy. We utilise transfer learning to optimise accuracy of just our question as well as to classify \emph{HST} data. Since this work, \texttt{Zoobot} has been trained on \emph{HST} data so the transfer learning step would not be needed in future with the new models. How we apply transfer learning is discussed in the following Section, but an excellent review and discussion of applying transfer learning for detecting galaxy mergers can be found in \citet{ackermann_18}.

\subsection{Transfer Learning}\label{trans_learn}
\noindent Transfer learning (or finetuning) is a method of applying the same machine learning model to a similar problem that it was originally trained on. Rather than having to completely retrain all parameters in a model and essentially create a new one, we can use the original model architecture and the parameters it has learned from its previous training. In the case of \texttt{Zoobot}, we keep the parameters it has learned from training on the DECaLS dataset and freeze all sections of the model responsible for feature extraction and recognition. 

We construct a classification section that maximises accuracy and only allow the weights of this section to change. As the classification section has fewer parameters than the feature extraction section (the classification section contains 86,209 parameters compared to the feature extraction sections' 4,048,989 parameters) we need significantly less data to completely retrain it (in our case, a factor of 15 less). Once this retraining is complete, the weights of the feature extraction sections of the model can be unfrozen and tweaked using our smaller dataset with a very low learning rate to further boost overall model accuracy.

An example of taking an existing model and applying it to a new problem with transfer learning is shown in \citet{walmsley22b}. Here, they take the trained model and finetune it to finding ring galaxies. They retain an accuracy of 89\% while only needing to train the model on 10$^{3}$ ring galaxies. This significantly reduces computational expense and training time of the model, while keeping the required training set very small. Interacting galaxies are rare, and interacting galaxy catalogues not expansive. So retraining the full network on hundreds of thousands of interacting galaxies is not feasible. Using transfer learning, and following the example from \citet{walmsley22b}, we only need to create a training set of 10$^{3}$ - 10$^{4}$ interacting galaxies to achieve an accuracy of $\approx$90\%. 


\section{CREATING THE TRAINING SET}\label{training}
\noindent We create a large training set of interacting galaxies following the criteria described in Section \ref{data} to train our model. Therefore, we need a large, labelled set of interacting and non-interacting galaxies. We elect to follow the methodology of finetuning as described in \citet{walmsley22b}, and aim to create a balanced training set. This has the advantage that it significantly improves the performance and accuracy of machine learning classifiers, but the disadvantage that it can bias our final model if few interacting galaxies exist compared to the general population. However, such a bias will be mitigated by using a high prediction cutoff to define an interacting galaxy. This is discussed in in Section \ref{performance}. To create this large training set we use the Galaxy Zoo collaboration \citep[initial data release described in][]{Lintott_08}.

\subsection{Interacting Galaxies and Galaxy Zoo}
\noindent The data in Galaxy Zoo is volunteer classifications on galaxy images spanning multiple projects. We incorporate classifications from all major Galaxy Zoo projects; Galaxy Zoo 1 \citep{Lintott_08}, Galaxy Zoo 2 \citep{gz:2}, Galaxy Zoo: \emph{Hubble} \citep{gz:h}, Galaxy Zoo: CANDELS \citep{gz:c} and Galaxy Zoo: DECaLS \citep{walmsley22}. These projects contain a total of 1,367,760 labelled galaxy images that we must extract the interacting galaxies from. We only use labels that are from citizen scientists, and no labels generated by previous versions of \texttt{Zoobot}. We apply three criteria to each interacting or non-interacting label. Firstly, it must have greater than 20 volunteer votes on it. Applying this allows us to use a statistically robust weighted vote from a crowd answer rather than trusting any volunteers individually. Secondly, the calculated weighted vote (i.e. the combination of the 20 or greater votes) must then be greater than 75\% in favour of being an interacting galaxy or less than or equal to 25\% for it not to be; this ensured purity in our training set. If the question given to volunteers was more specific (such as `Is this a minor disturbance?' and `Is this a major disturbance?') then if either answer was the majority vote we classified it as an interacting galaxy. Thirdly, the object must exist in the \emph{Hubble} footprint so that we could make a cutout of it.

Checking if each training source existed in the \emph{Hubble} footprint was only possible in an efficient way because of ESA Datalabs. Rather than having querying every coordinate and make network calls to TAP services, we extract every final product $F814W$ observation footprint and check if each labelled galaxy exists in at least one file. We make this check by creating a \texttt{Shapely} Polygon for each observational footprint and a \texttt{Shapely} Point for each labelled galaxy central coordinate. Using the \texttt{Shapely} Polygon \textsc{contains} function, we check if a labelled galaxy's Point overlaps with an observations' footprint Polygon. This returns a list of files which contain the training source. If a training source was not found in any observational footprint we discard it. We make no attempt here to check if our sources have other photometry available to them, and only create 1-color images with the $F814W$ data. We provide the images to \texttt{Zoobot} as 1-color grayscaled cutouts.

Upon applying these criteria we find 3,167 labelled interacting galaxies in Galaxy Zoo: \emph{Hubble} project, the largest contribution to our training set. These were paired with 3,167 labelled non-interacting systems (following the previous criteria) to balance the training set. From all other projects, we find 869 labelled interacting systems which fitting the creation criteria. The primary limiting factor for Galaxy Zoo's 1 and 2 was that many found interacting galaxies did not exist in the Hubble footprint. For Galaxy Zoo: CANDELS and Galaxy Zoo: DECaLS the limiting factor was the required calculated weighted vote. These labelled interacting systems were then paired with 869 labelled non-interacting systems, ensuring that each labelled non-interacting system came from the same project as its labelled interacting system counterpart.

Each of these projects has a varied redshift range: Galaxy Zoo: \emph{Hubble} is $z < 1$, Galaxy Zoo: CANDELS $1 < \emph{z} < 3$ and Galaxy Zoo's 1, 2 and DECaLs are $\emph{z} < 0.15$. This introduces a redshift bias into our model, where the morphology and brightness of interacting sources changes with a $z > 1$. This is only partially rectified by including Galaxy Zoo: CANDELS, which provided 322 labelled interacting systems.

\begin{figure*}
  \centering
  \includegraphics[width=0.85\textwidth]{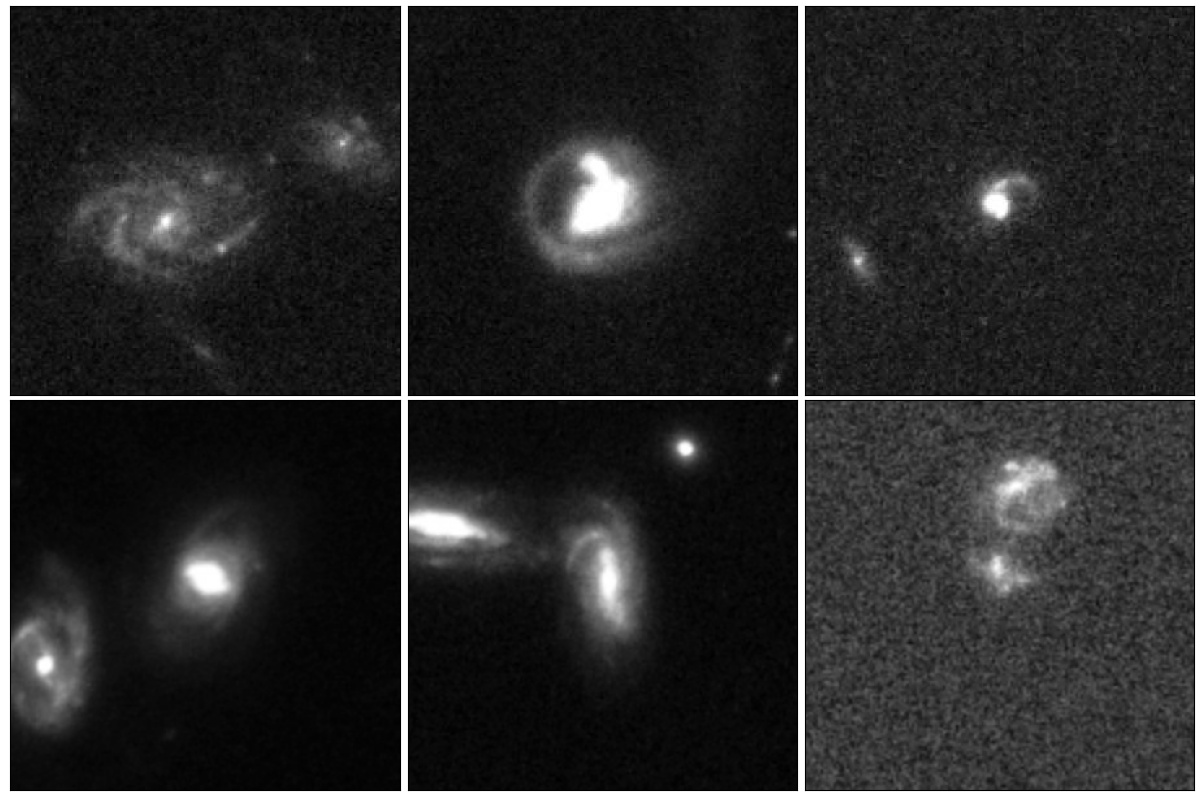}
  \caption{Example images of the labelled interacting galaxy systems used to train \texttt{Zoobot}. Each galaxy had a weighted vote fraction $\geq$0.75 in Galaxy Zoo. \textit{Top Row}: Three examples from the Galaxy Zoo: \emph{Hubble} project of the training set. \textit{Bottom Row}: Three examples from the other Galaxy Zoo projects. These are, from right to left, Galaxy Zoo 2, Galaxy Zoo CANDELS and Galaxy Zoo DECaLS. The priority with this training set was that the interactors had clear tidal features and disruption so \texttt{Zoobot} would learn to highly weight them and not misclassify close pairs.}
  \label{fig:training_set_int_ex}
\end{figure*}

\begin{figure*}
  \centering
  \includegraphics[width=0.85\textwidth]{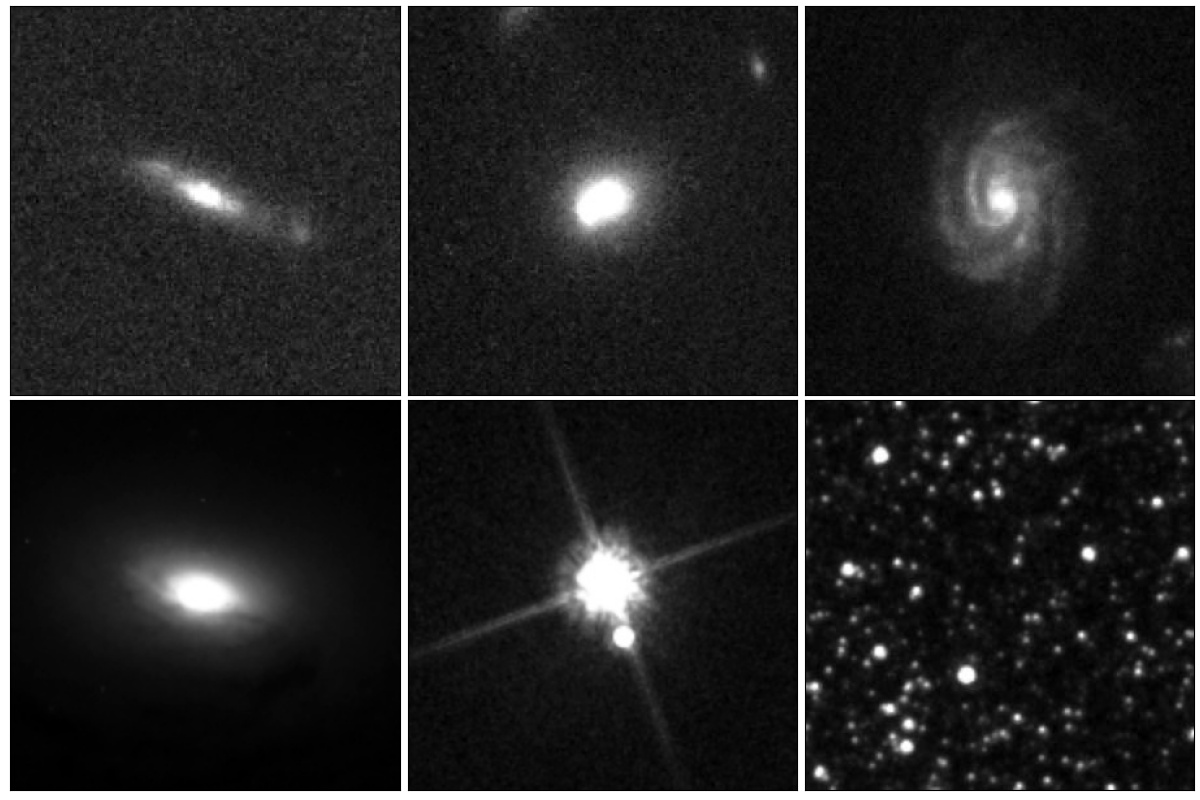}
  \caption{Example images of the labelled non-interacting galaxy systems used to train \texttt{Zoobot}. \textit{Top Row}: Three examples from the Galaxy Zoo: \emph{Hubble} project of the training set. \textit{Bottom Row}: Three examples from the other Galaxy Zoo projects. These are, from right to left, Galaxy Zoo 2, Galaxy Zoo CANDELS and a starfield from the active learning cycle. Starfields/globular clusters/open clusters existed throughout the HSC flagged as extended sources. 1,000 images of starfields were added to the training set so \texttt{Zoobot} would give them a very low score.}
  \label{fig:training_set_non_ex}
\end{figure*}

From all Galaxy Zoo projects, we find a training set of 4,036 labelled interacting galaxies and combine them with their matched 4,036 labelled non-interacting galaxies giving a total training set size of 8,072. Figures \ref{fig:training_set_int_ex} and \ref{fig:training_set_non_ex} show six examples of our labelled interacting and non-interacting galaxy training set. As we require \texttt{Zoobot} to learn to weight tidal features or disturbances highly, it is important that such structures dominate the training set. Previous works, such as \citet{Pearson_22}, have found that final catalogues produced by CNNs are often heavily contaminated by sources which are simply close pairs by projection effects and chance alignment in the sky. By focusing our CNN on tidal features, we aim to minimise this contamination. We ran an initial test of the prediction pipeline on the first 500,000 sources that had been created from the HSC to initially test our \texttt{Zoobot} model. We investigate any source which was given a prediction score $\geq 0.75$ and, to further increase the size of our training set, conduct one step of active learning.

\subsection{One Active Learning Cycle}
\noindent To enlarge our training set further, we conduct one step of active learning to find interacting galaxies. An active learning cycle involves an `expert' checking the predictions made by the model, correcting any incorrect predictions and then feeding it back into the model as additional labelled images to a training set. We complete finetuning of \texttt{Zoobot} on our initial training set of 8,072 galaxies and make predictions on the first 500,000 sources from the HSC (created under the criteria previously discussed). We visually inspect the sources \texttt{Zoobot} gives a prediction score $\geq0.75$ and correct any wrong predictions. These corrected labelled sources and those \texttt{Zoobot} correctly labelled are then added to the training set. Not only does this step allow us to add more labelled interacting galaxies to the training set, but it also allows us to evaluate \texttt{Zoobot}'s behaviour and check if it consistently predicts a type of source or galactic morphology incorrectly.

From the first 500,000 sources, a total of 6,198 sources were given a prediction score of $\geq$0.75. We correct the predictions \texttt{Zoobot} made and balance this set to 5,698. During this cycle, a large number of globular clusters/starfields/open clusters were given a very high prediction score. Figure \ref{fig:training_set_non_ex} shows an example of these contaminating star fields. We created sources of 1,250 star fields and added these into the training set, labelling them as non-interacting. Adding the balanced 5,698 sources plus the 1,250 starfields to our training set gave us an unbalanced training set of 15,020 sources. To then balance the training set, we took 1,250 labelled interacting galaxies from the Galaxy Zoo: \emph{Hubble} project and made random image augmentations with the \texttt{TensorFlow} Python package. These augmentations were simple rotations, cropping and resizing. With these extra sources, our training set contains 16,270 sources. Of these, 50\% (8,135) were labelled images of interacting galaxy systems.


\section{DIAGNOSTICS}\label{diagnostics}
\subsection{Model Performance} \label{performance}
\noindent Upon finetuning \texttt{Zoobot} we validate its performance. We reuse the validation set that \texttt{Zoobot} automatically creates when training. This set is created by putting aside a random set of 20\% of the training set. \texttt{Zoobots} then uses it to validate its performance in training. We record which images \texttt{Zoobot} selected, and extract these from the training set for further diagnostics. This provides us with a validation set of 3,270 images, containing 1,648 non-interacting galaxies and 1,622 interacting galaxies.

\begin{figure}
  \centering
  \includegraphics[width=\columnwidth]{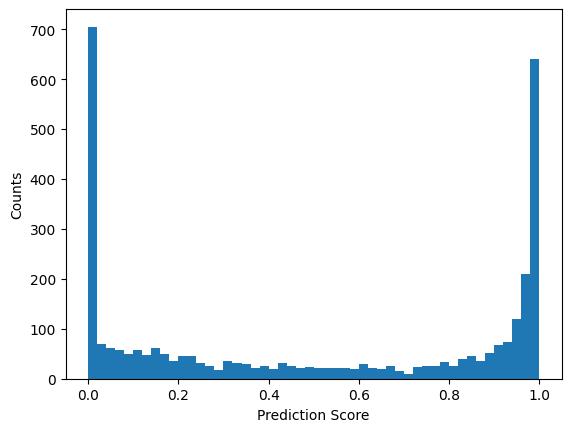}
  \caption{The distribution of prediction scores given to our validation set of 3,270 labelled sources set aside by \texttt{Zoobot} in training. These were split into 1,648 non-interacting sources and 1,622 interacting sources. As can be seen from the distribution, our model is often confident when a source does or does not contain an interacting galaxy by the strong bi-modality. This is likely due to the very stringent vote weightings used when selecting the training set. Using this distribution, we decide the prediction score to use as a cutoff to give us our final binary classification: interacting galaxy or not.}
  \label{fig:pred-score}
\end{figure}

\texttt{Zoobot} gave a prediction score between 0 and 1 to each of the validation images, Figure \ref{fig:pred-score} shows the resulting distribution. This distribution shows that our model has high confidence in what is or isn't an interacting system due to the high counts at very low and very high probability scores. It is likely the use of a balanced training set, and the very low volunteer score needed to define a source as non-interacting that leads to a strongly bi-model prediction score distribution. Using a balanced training set is an intrinsic trade off between ease of training, and potential biases introduced. Having a balanced dataset does not reflect reality, and leads \texttt{Zoobot} to over-predict interacting galaxies. Using very stringent volunteer classification cutoffs also leaves few ambiguous systems in the validation set, further enhancing this bi-modality.

\begin{figure}
  \centering
  \includegraphics[width=\columnwidth]{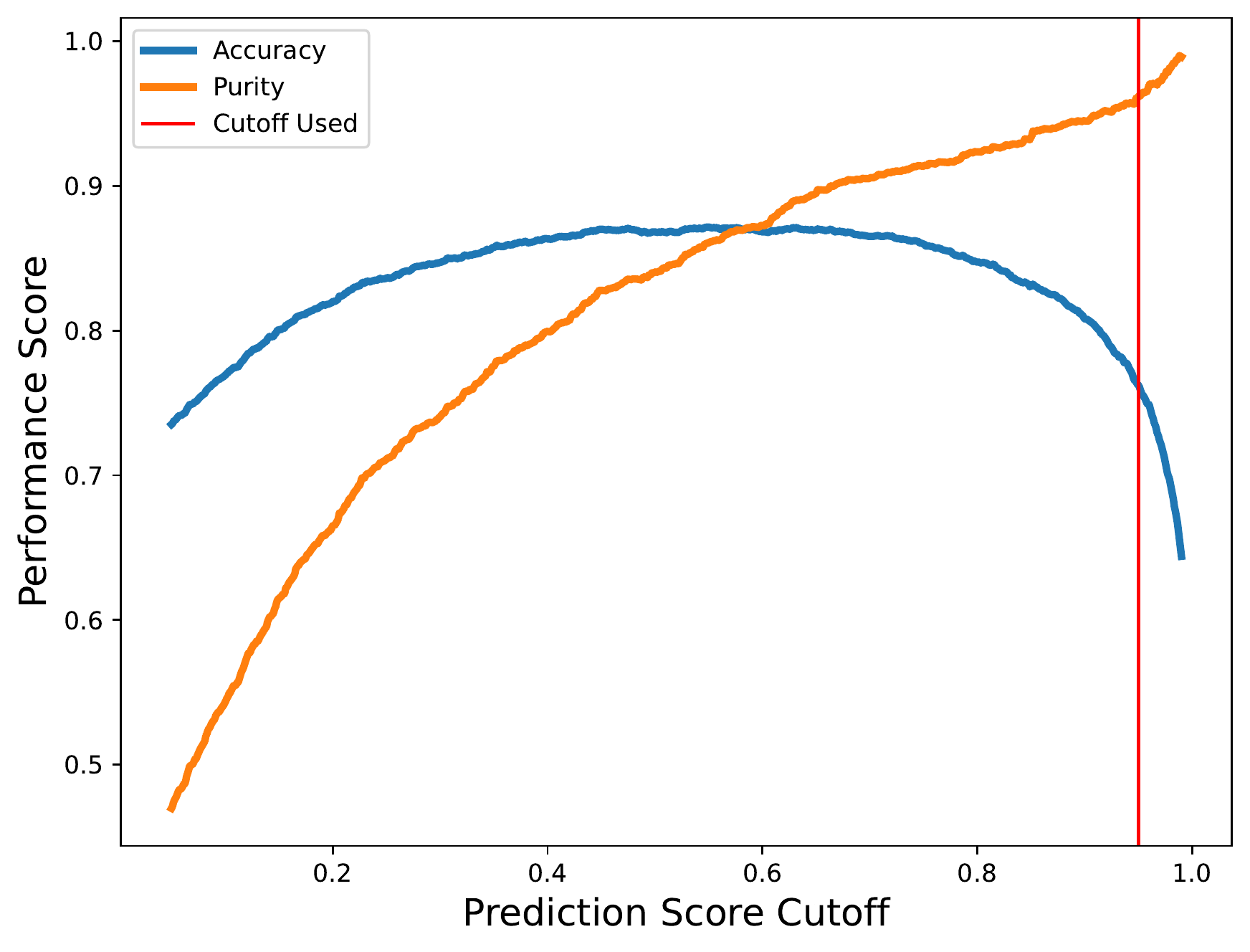}
  \caption{A measure of accuracy and purity against prediction score. The accuracy (in blue) is a direct measure of the number of sources \texttt{Zoobot} correctly predicted vs the total number of predictions made. The measure of purity (in orange) is the the number of predictions \texttt{Zoobot} correctly made vs the total number of predictions for an interacting galaxy. The cutoff score (in red) shows the point above which we would define an interacting galaxy and below which we would not. At this point, the accuracy appears lower due to \texttt{Zoobot} making many false negative predictions while successfully making true negative predictions. This is confirmed by the maximisation of purity. Due to the number of sources \texttt{Zoobot} is predicting over, the size of the catalogue will exceed any previous catalogues. Therefore, we use this very conservative cutoff to maximise purity over the completeness of our catalogue. These measures can also be shown with the F1 score. Figure \ref{fig:f1-score} shows this change with prediction cutoff in the Appendix.}
  \label{fig:accuracy-graph}
\end{figure}

The prediction score must be reduced to a binary classification for our problem. We use Figure \ref{fig:pred-score} to define a prediction score above which a source is classified as an interacting galaxy. We measure the accuracy of \texttt{Zoobot} for different cutoffs, where the accuracy is the fraction of labels correctly predicted over the total number of labels predicted on. Figure \ref{fig:accuracy-graph} shows this change in accuracy. We find that our model is most accurate with a prediction score cutoff of 0.55 with an accuracy of 88.2\%. Figure \ref{fig:accuracy-graph} also shows the change in the purity of our catalogue with changing prediction cutoff. Here, purity is the ratio of number of true interacting galaxies to total sources in the final catalogue. These scores can be combined into the F1 score of our model, shown in Figure \ref{fig:f1-score} in the Appendix.

\begin{figure}
  \centering
  \includegraphics[width=\columnwidth]{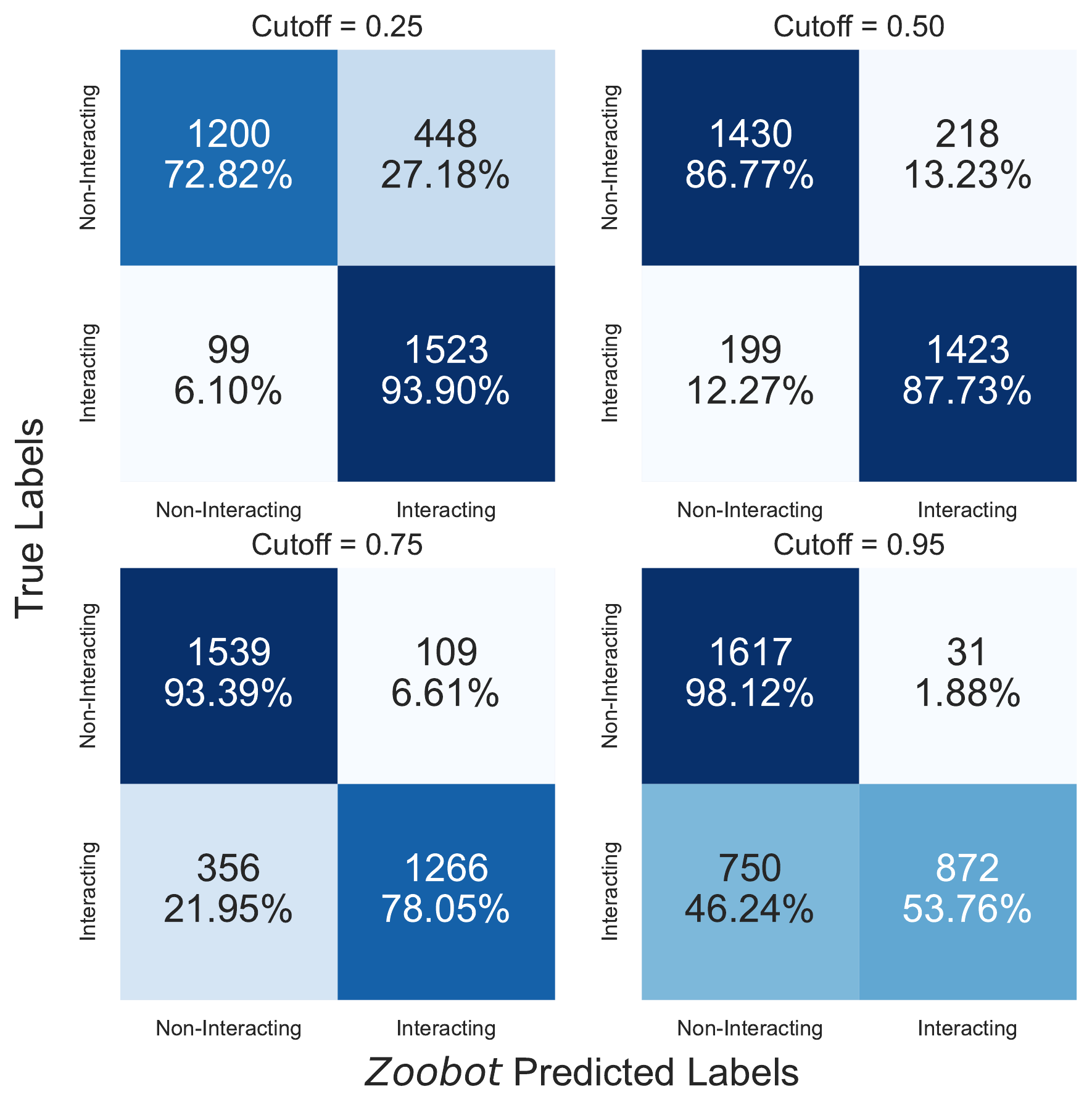}
  \caption{Confusion matrices of four different cutoffs of prediction score defining a binary classification of interacting galaxy or not. Confusion matrices break down our accuracy measurement into how \texttt{Zoobot} is misclassifying sources. At a cutoff of 0.50, the accuracy is highest at 88.2\%. However, at this cutoff, $\approx$10\% of our final catalogue would contain contamination. We elect to use the very stringent prediction cutoff of 0.95 for the rest of this work as it will return the lowest contamination.}
  \label{fig:conf-mat}
\end{figure}

Figure \ref{fig:conf-mat} also shows a measure of accuracy for our model at different cutoffs using confusion matrices. Importantly, it also shows how our model is getting labels wrong: either giving false positives (where a labelled non-interacting galaxy is predicted to be interacting) or false negatives (where a labelled interacting galaxy is predicted to be a non-interacting). The number of incorrect positive and negative predictions change based on the prediction cutoff, with a very low cutoff giving many false positives and a very high cutoff giving many false negatives. Figure \ref{fig:conf-mat} shows that with a cutoff of 0.50, we would return a high level contamination in our final catalogue. Of the 1,622 galaxies predicted to be interacting, 218 would be non-interacting systems - approximately 13\%. Our main aim in this work is to present a highly pure, large interacting galaxy catalogue that can be used for statistical exploration of interacting galaxy parameter space. Therefore, we use a very stringent cutoff of 0.95.

Using a cutoff of 0.95 reduces contamination significantly. Figure \ref{fig:conf-mat} shows the final contamination in our validation catalogue would be $\approx$2\%, where Figure \ref{fig:accuracy-graph} shows that we are maximising the purity in our sample at the expense of accuracy. The aim of this work is not to create a general tool to be used by the community, but to find a large catalogue of interacting galaxies. As we are investigating 126 million sources, despite removing $\approx$50\% of interacting galaxies from the final catalogue, we are certain that we can find a catalogue larger than previous works.

Using such a high cutoff also reduces any risk of any biases introduced by using a balanced training set. While using such a training set often increases the accuracy and speeds up training, it can bias the model towards one conclusion. In our case, the true rate of interacting galaxies will be much smaller than 50\%. Therefore, our model will be biased to labelling a source as an interacting galaxy. This will be particularly true for edge cases, which could be ambiguous to even an expert classifier. By using such a high cutoff score, this bias will be mitigated by only labelling the most clearly interacting objects as interacting.

\begin{figure}
    \centering
    \includegraphics[width=\columnwidth]{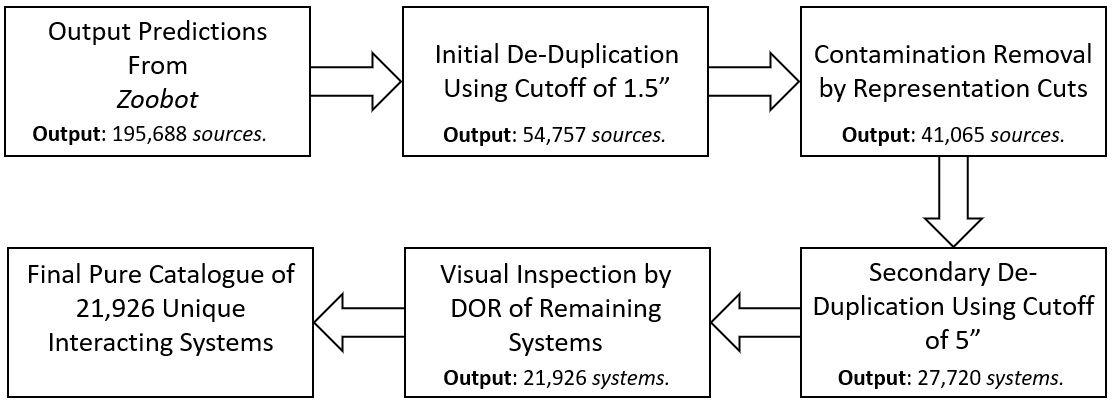}
    \caption{Flow diagram of our contamination and duplication removal process. De-duplication used agglomerative clustering based on sky separation. The first step of de-duplication uses a cutoff of 1.5$^{\prime\prime}$. This significantly reduced duplication in the catalogue, as well as the size of the catalogue to 54,757 interacting galaxies. We then applied contamination removal to this de-duplicated catalogue. Upon visual inspection, a small number of duplicated systems still existed in the catalogue. To ensure a pure catalogue of unique systems, we applied a agglomerative clustering again with a cutoff of 5$^{\prime\prime}$. This gave us a catalogue of 27,720 unique interacting systems. The final step to ensure purity was visual inspection by DOR, removing any remaining contamination. This gave the final pure catalogue of 21,926 unique interacting systems.}
    \label{fig:flow-diagram}
\end{figure}

\subsection{Duplication Removal}\label{dup_rem}
\noindent The fully trained \texttt{Zoobot} made predictions on $\approx$126 million extended sources from the HSC that had passed our creation criteria. Of these, 195,688 sources were given a score of 0.95 or greater, $\approx$0.2\% of the total number of sources. Upon visually inspecting a subset of sources, it is clear that our \texttt{Zoobot} model had predicted for an interacting galaxy even if it was not the central (and, therefore, target) source in the image. This is due to the misalignment of sources from the centre in the training set as described in Section \ref{training}. \texttt{Zoobot} learned to classify an image as an interacting galaxy if it contained one, and not just if it was the central source. Therefore, many interacting systems were duplicated in our final catalogue, appearing in cutouts were the central source was not interacting.

Another source of further duplication was the HSC itself. In the HSC, many extended objects have multiple source IDs applied to them. This is due to bright clumps in extended sources being assigned a new ID, sources which had been found but did not exist in reality or background sources which existed in extended systems. We find that of the 195,688 Source IDs given a prediction score of 0.95 or greater, approximately 3.6 Source IDs were matched to a single real object. To refine the catalogue and remove the duplication we use spatial clustering of each source with agglomerative clustering \citep[an introduction and description of hierarchical clustering, including agglomerative clustering, can be found in][]{nielsen16}.

Agglomerative clustering is a method of hierarchical clustering based on a distance metric between the sources. We set the maximum distance between points to define a cluster. i.e. any sources within a defined distance on the sky from each other will be merged under one source ID. This approach means we do not need any knowledge of how many cluster of sources exist in the dataset or the level of duplication within it, as would be the case in many other clustering approaches. We create distance matrices of the angular separation of every source using the \texttt{Astropy} Python package. These projected sky separations are then used as a euclidean distance in the clustering algorithm with an \textsc{euclidean\_linkage}. The new ID of a cluster is the first source ID in the cluster. 

Initially, we utilise a limiting sky separation of 1.5$^{\prime\prime}$ to remove the duplication. This reduced the size of our potential catalogue to 54,757 interacting galaxy candidates. We then applied contamination removal as described in Section \ref{bad_pred}. Once contamination removal was completed, the catalogue size was 41,065 interacting galaxies. Visual inspection found further duplication, so our initial de-duplication had not been aggressive enough. To ensure the catalogue was of unique systems, we opted to use a final aggressive limiting sky separation of 5$^{\prime\prime}$ completely removing the duplication in our catalogue. This aggressive de-duplication further reduced the size of our catalogue to 27,720 candidate interacting systems. However, we could be certain that each of these candidate systems was unique. Figure \ref{fig:flow-diagram} shows a full breakdown of the steps in our de-duplication and contamination removal process.

\subsection{Bad Predictions \& Removal}\label{bad_pred}
\noindent After the initial step of de-duplication we begin removal of contamination from the catalogue. A major, and expected, source of contamination is by close pairs of galaxies. These are systems where chance alignment in the sky appears that galaxies are close together but are actually at different redshifts. Other sources of contamination include large central galaxies with satellite galaxies about them, star fields with extended sources in them and objects with strange morphologies that \texttt{Zoobot} predicted were tidal features.

Upon applying the clustering by sky projection of 1.5$^{\prime\prime}$, the catalogue contained 54,757 candidate interacting galaxies. Our primary concern is contamination by close pairs. Creating catalogues of interacting galaxies with CNNs are notorious for suffering from this problem, where a significant number of candidates must be removed from otherwise large final catalogues \citep{bottrell_19, Pearson_22}. The decisive way to remove this contamination is to compare redshift measurements of each galaxy in the candidate interacting system. However, this is impractical for our catalogue where the majority of candidates have no redshift measurements. To find close pairs, and remove them effectively, we take advantage of the representations \texttt{Zoobot} learns of each image. As described previously, \texttt{Zoobot} was trained to answer every question in Galaxy Zoo: DECaLS simultaneously for every galaxy. It therefore learns a generaliseable representation of many kinds of galaxies. In this representation space, morphologically similar galaxies will exist close together in clusters while those that are dissimilar will be further apart. We extract the features \texttt{Zoobot} has learned of each candidate, and plot its representation.

We remove the classification head of \texttt{Zoobot} and directly output the final layer of the feature learning section of the model. This gives 1,280 features (the representations) for each of our 27,720 candidate systems. However, there will be much redundant information in this very high dimensional feature space. We compress this using incremental principal component analysis (PCA) \citep{ross08}. An excellent demonstration of using this approach can be found in \citet{walmsley22b}. We reduce the dimensionality from 1,280 to 40{(as in \citet{walmsley22b}), and input the resultant components into the Auto-Encoder \texttt{UMAP} \citep{umap}. \texttt{UMAP} projects the 40 dimensional components of each candidate system onto a 2 dimensional manifold. The position of each galaxy on this manifold is directly linked to its visual morphology. Close pairs have similar visual features which will then appear as a cluster in our representation space.

\begin{figure*}
  \centering
  \includegraphics[width=\textwidth]{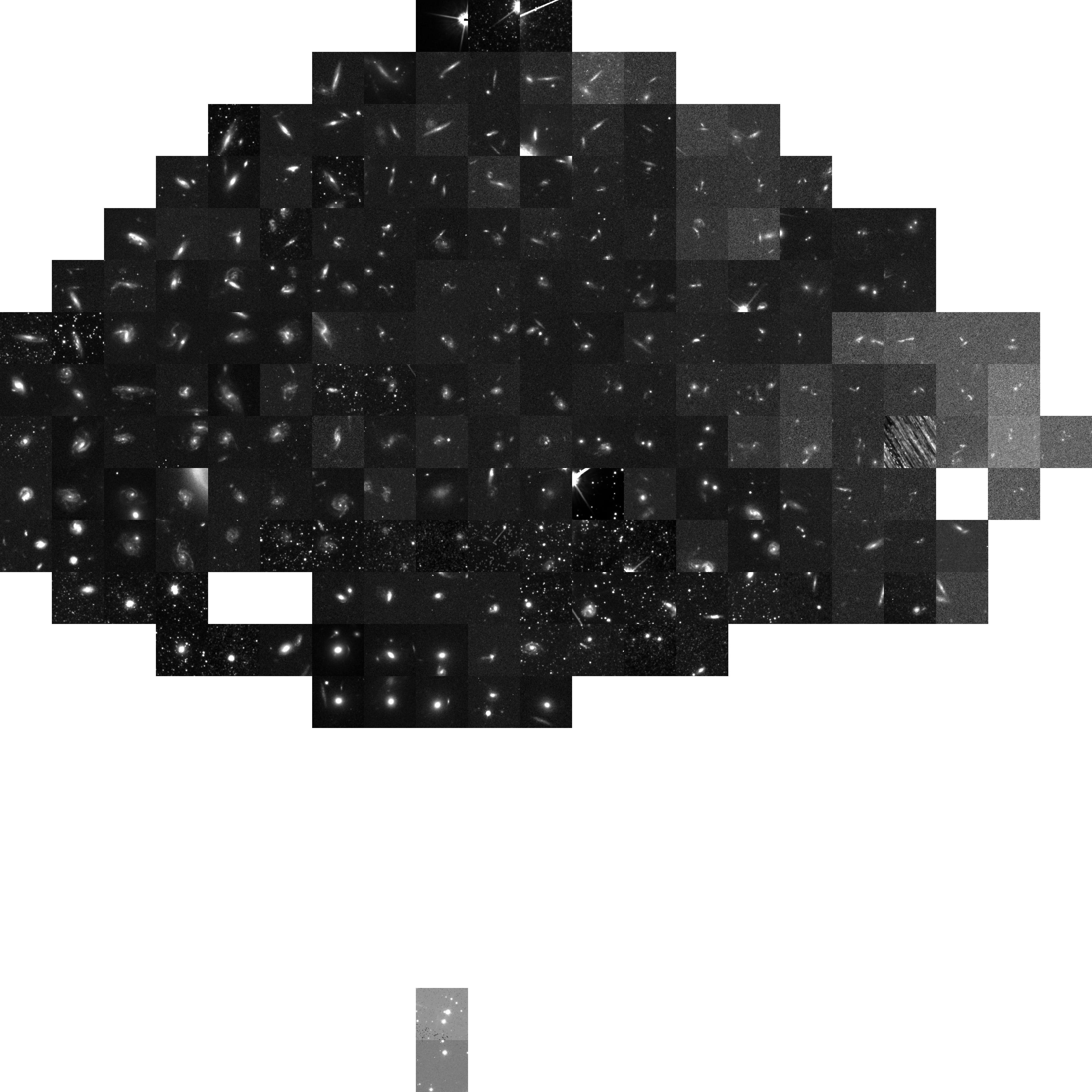}
  \caption{The representation distribution of 54,757 candidate interacting galaxies. This distribution is the compressed 2D representation of the 1,280 dimensional representation that \texttt{Zoobot} has learned of each image. Each image is a randomly selected one from sources within each bin in the distribution. The X and Y axis on this plot are the 2D mapping on the manifold given by UMAP for the 40 dimensional principal components of each source, and not physical parameters. Three gradients are clear in this distribution: first; from the left to right there is a distinct gradient in the contrast of the images. The images to the left are local galaxies with low redshift, while those on the right are dimmer sources at much higher redshift. This is an effect of how the images are created using a linear scaling function and a fixed contrast. The second feature, also from left to right, is a gradient of larger source size to smaller source size. This is a feature \texttt{Zoobot} has learned based on the redshift of the source as well. The third, from top to bottom, is a gradient of the inclination of the source. With the most inclined (and even diffraction spikes) of the sources appearing at the top, while at the bottom the sources are face on. Along the bottom of the representation plot, there are close paired sources as well as many star fields. Along the very top, there is contamination in the form of isolated stars in star fields. Thus, we make aggressive cuts along the top and bottom of our representation space to remove as much contamination in a general way. The full representation plot, with all sources and the cuts, is shown in Figure \ref{fig:cuts-visual}. }
  \label{fig:representations}
\end{figure*}

\begin{figure}
  \centering
  \includegraphics[width=\columnwidth]{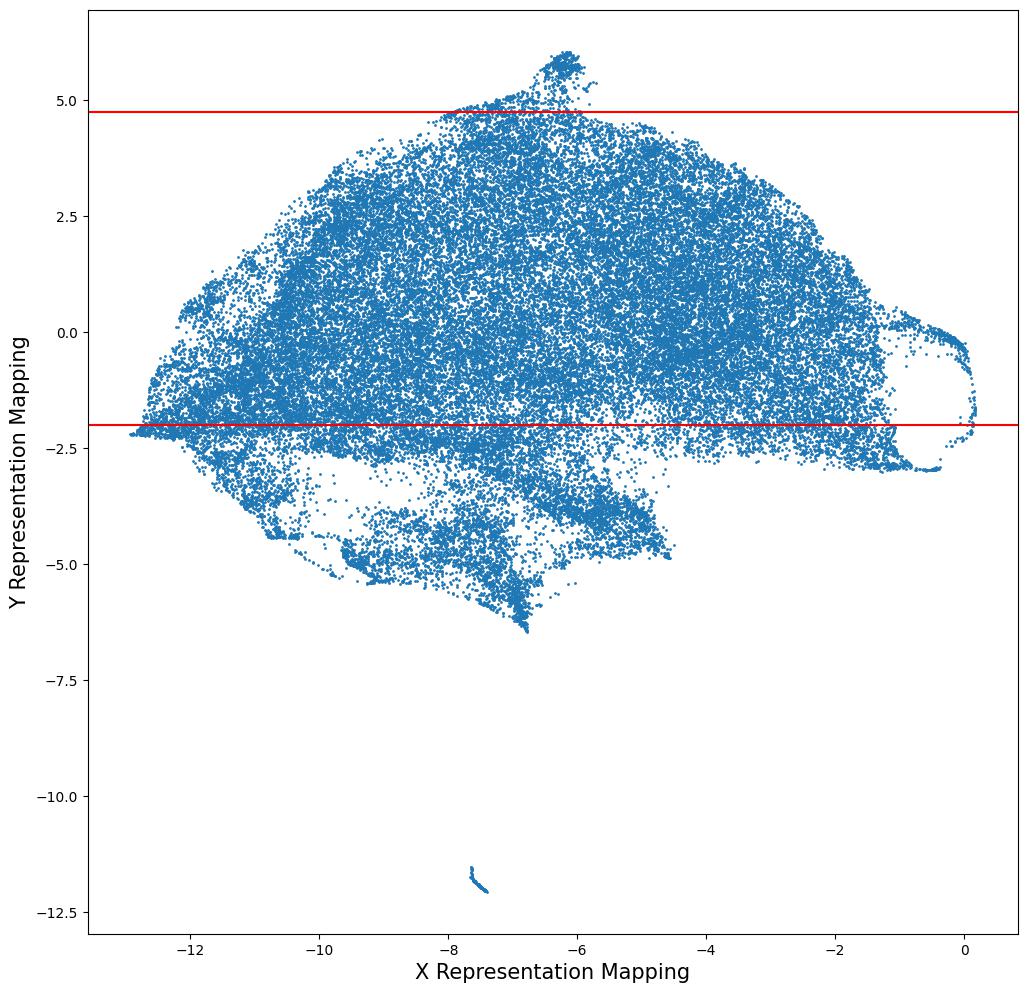}
  \caption{Scatter plot showing the precise distribution of each representation of sources in the remaining 54,757 sources. This is the unbinned version of Figure \ref{fig:representations}. The two red lines show the cutoffs utilised to remove the majority of close pairs by projection as well as the very obvious contamination of stars and stellar fields at the top of the representation distribution. The number of candidate interacting systems in the catalogue was reduced to 41,065 systems.}
  \label{fig:cuts-visual}
\end{figure}

Figure \ref{fig:representations} shows the representation distribution of our 54,757 candidates after compression with \texttt{UMAP}. A random image in each bin has been selected to show the morphology of the objects within the bin. There are three clear gradients that exist in the representation distribution: one of source size, one of the source inclination and one of image contrast between the source and the background. The gradient of source size is clear from left to right. This is also true of contrast between the source and background. The gradient of source inclination is from top to bottom. The top shows very inclined sources, and even the diffraction spikes of stars, while along the bottom we find face on sources which take up a larger part of the cutout centre. At the very bottom of the figure (away from the main body) a cluster of very poorly contrasted sources with the background that are face on are found. The gradients of inclination and source size are expected while that of contrast is less so. This gradient is likely a result of how we created our images using a Linear Stretch with fixed contrast. The effect of this is that dimmer sources have brighter backgrounds, a particular issue at high redshift.

Figure \ref{fig:representations} has many areas of similar morphology. On the left, we have isolated objects: disturbed spirals or large galaxies with tidal disturbance to them. Along the bottom, we see isolated bright objects with satellites about them. On the bottom right, we see our area of representation space dominated by close pairs. In the centre, we see the population of interacting galaxies that \texttt{Zoobot} was trained to find. The areas of representation space which are dominated by clear sources of contamination are cut. Figure \ref{fig:cuts-visual} shows a scatter plot of the representation distribution and the cuts we make. They are made such that any source with a Y Mapping of $-2 \leq Y \leq 4.75$ will be kept in the catalogue. The choice of these cuts has been made by eye, and then bootstrapping the remaining images to check contamination removed. After applying these cuts, we retain 41,065 systems in our catalogue.

We estimate $\approx$ 25\% of sources in the greater than 0.95 prediction bin are close pairs. This may seem lower than previous works, but is due to our very conservative prediction cutoff. The general cuts to our population based on their position in representation space makes it very likely that we retain some close pairs in the catalogue, while also removing interacting galaxy systems.

As described in Section \ref{dup_rem}, we then apply a 5$^{\prime \prime}$ to the 41,065 remaining candidates, further reducing our catalogue to 27,720 systems. With such an aggressive sky projection cut, many individual interacting galaxies are now identified under the same ID as the secondary galaxy in the system. To remove remaining contamination in the catalogue, a final visual classification step was conducted. This visual inspection was conducted by DOR. Any systems removed at this stage were classified into three categories: interacting system, contamination and gems. The gems sub-category became necessary as many sources of contamination that were being removed were objects of other astrophysical interest, and is described in Section \ref{gems}.

\section{RESULTS \& DISCUSSION}\label{results}
\subsection{An Interacting Galaxy Catalogue}\label{int_cat}
\noindent Upon de-duplication and contamination removal described in Sections \ref{dup_rem} and \ref{bad_pred}, our final catalogue contains 21,926 interacting systems. Figure \ref{fig:interactors} shows a random sample of 50 of the systems from our catalogue. In these examples we can see highly distorted or currently interacting systems, precisely what we trained \texttt{Zoobot} to highly predict. Some cutouts are of the full interacting system, containing both the primary and secondary galaxies in the interaction. Some source cutouts only show one of the interacting galaxies, though these systems remain highly disturbed. Due to the constraints in our training set, so highly weighting disturbance or tidal features in our predictions, we are sampling interaction from all epochs except the approach to the initial pass. At this initial stage, there will be no tidal features formed or disturbance in the disks as the two galaxies approach each other. Separating them from close pairs would be difficult without kinematic or redshift information, not available for the majority of these sources.

\begin{figure*}
  \centering
  \includegraphics[width=0.80\textwidth]{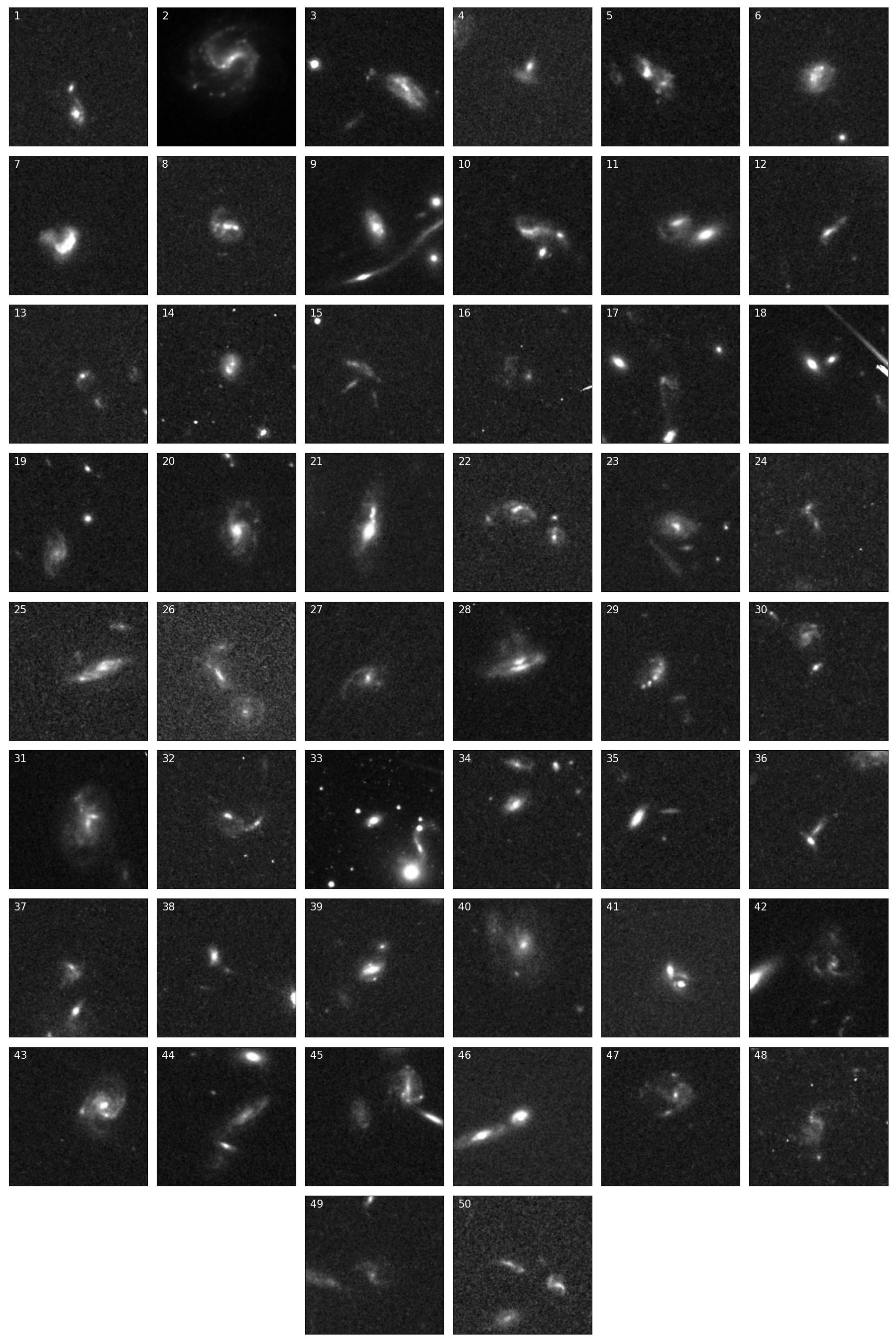}
  \caption{An example of 50 of the final interacting systems found with \texttt{Zoobot}. These were selected randomly from the de-duplicated and de-contaminated 21,926 sources. Each of these examples have extended tidal features and distortion. Not all of the final interacting systems have two galaxies within them (for example, image 2), but are clearly very disturbed by a tidal event. These were kept in as they would form a large part of the interacting galaxy population and would be flagged as disturbed or interacting in Galaxy Zoo. Each of these images is a 1-colour image using the $F814W$ \emph{HST} filter.}
  \label{fig:interactors}
\end{figure*}

We investigate which of the systems in our catalogue have previous references in the astrophysical literature. To search the literature, we use the \texttt{AstroQuery} Python package with a coordinates based search of cutoff radius 5$^{\prime\prime}$. We search the astronomical databases Simbad, the NASA Extragalactic Database (NED) and ViZier for references to our interacting systems. These return either a list of references, or an empty list showing no references associated with the system. We find that 7,522 of our systems have at least 1 reference associated with them, while and 14,404 do not. A flag exists in the catalogue data release which shows whether a system has references associated with it or it could be considered a `new' system. We, however, do not claim that these systems are discovered by ourselves. These systems have always existed in the backgrounds of large surveys or observations and been discovered by others, it is only with ESA Datalabs that we can apply a methodology such as in this work to extract those systems from these observations. We also do not claim that these unreferenced systems are particularly interesting or phenomenal. It is most likely that these systems are the very faint background galaxies in surveys or observations whose main objective was something other than finding interacting galaxies. This will be further discussed in Section \ref{z_phot_analysis}.

Figure \ref{fig:sky-distribution} shows the distribution of our catalogue in the sky. The \emph{HST} is able to observe the majority so the catalogue sources are scattered throughout it. We find that the sources cluster in different parts of the sky which correspond to major surveys conducted using the \emph{HST} involving ACS/WFC and the $F814W$ filter. We also mark the centres of the seven surveys which correspond to the major clustering of interacting systems in the sky. These were the COSMOS, the GOODS North, GOODS South, PHAT, CANDELS, AEGIS and Spitzer Space Telescope FLSv Region \citep{morganti04} surveys.

The full catalogue and data product are found on Zenodo at the following DOI where it is freely accessible to the community: \url{https://doi.org/10.5281/zenodo.7684876}. Table \ref{tab:ex-cat} shows an example of the data and format of the 50 sources shown in Figure \ref{fig:interactors}. We also bootstrap the final catalogue as an estimate of contamination remaining. As described in Section \ref{bad_pred}, the final step of contamination removal was visual inspection by DOR of the 27,720 candidate interacting systems to remove the remaining 5,794 contaminants from the final catalogue. Visual inspection by a single expert at this scale is not perfect. We extract random sources from the catalogue in batches of 500 and manually re-classify them again. This bootstrapping reveals that $\approx$3\% of our interacting system in the final catalogue remains contamination.

\begin{deluxetable*}{ccccccc}
  \tabletypesize{\scriptsize}
  \tablewidth{0pt}
  \tablecaption{An example of the format of the final catalogue for the 50 example images presented in this paper. \label{tab:ex-cat}}
  \tablehead{
  \colhead{Image No.} & \colhead{SourceID} & \colhead{RA (deg)} & \colhead{Dec (deg)} & \colhead{Interaction Prediction} & \colhead{References} & \colhead{Status}
  }
  \colnumbers
  \startdata
  1 & 4001014298177 & 261.292845 & 37.162387 & 0.983999 & No entry &  Unreferenced \\
  2 & 4001444190958 & 183.527536 & 33.183451 & 0.998016 & [1994PASP..106..646K] & Referenced \\
  3 & 4000809226818 & 93.960150 & -57.813401 & 0.982266 & [2019ApJ...878...66C] & Referenced \\
  4 & 4553390202 & 73.581297 & 2.903528 & 0.968280 & No entry & Unreferenced \\
  5 & 4000907600174 & 259.037474 & 59.657617 & 0.999978 & No entry &  Unreferenced \\
  6 & 4575187799 & 150.001883 & 2.731942 & 0.974649 & [2007ApJS..172...99C] & Referenced \\
  7 & 4000717342023 & 149.527791 & 2.126945 & 0.993912 & [2007ApJS..172...99C] & Referenced \\
  8 & 4001174802281 & 28.593114 & -59.643515 & 0.982890 & No entry & Unreferenced \\
  9 & 4182689774 & 186.709991 & 21.835419 & 0.973232 & [2016ApJS..224....1R, 2011ApJS..193....8B] & Referenced \\
  10 & 4000958398690 & 186.719496 & 23.961225 & 0.999288 & No entry & Unreferenced \\
  11 & 4266881925 & 344.730228 & -34.799824 & 1.000000 & No entry & Unreferenced \\
  12 & 4001084105393 & 150.128198 & 2.623949 & 0.982739 & [2018ApJ...858...77H, 2007ApJS..172...99C] & Referenced \\
  13 & 4000961670486 & 345.337556 & -38.985521 & 0.954961 & No entry & Unreferenced \\
  14 & 4000719687395 & 338.173538 & 31.189718 & 0.974724 & No entry & Unreferenced \\
  15 & 4001435343326 & 331.771500 & -27.826175 & 0.986885 & No entry & Unreferenced \\
  16 & 4001268932937 & 8.856781 & -20.271978 & 0.986329 & No entry & Unreferenced \\
  17 & 4651336656 & 149.836709 & 2.141702 & 0.984389 & [2007ApJS..172...99C] & Referenced \\
  18 & 4000877021787 & 116.211231 & 39.462563 & 0.979178 & No entry & Unreferenced \\
  19 & 4000878525229 & 149.834893 & 2.516816 & 0.963694 & [2007ApJS..172...99C, 2009ApJS..184..218L] & Referenced \\
  20 & 6000290755870 & 186.774907 & 23.866311 & 0.981961 & No entry & Unreferenced \\
  21 & 4000806637434 & 210.253419 & 2.854869 & 0.960790 & No entry & Unreferenced \\
  22 & 4001215753971 & 135.898809 & 50.487130 & 0.998386 & No entry & Unreferenced \\
  23 & 4000813961830 & 163.678042 & -12.776815 & 0.958405 & [2005ApJ...630..206F] & Referenced \\
  24 & 4001200639012 & 54.037618 & -45.170026 & 0.991404 & No entry & Unreferenced \\
  25 & 4000921402261 & 150.417634 & 2.313781 & 0.990775 & [2018ApJ...858...77H, 2012ApJ...753..121K] & Referenced \\
  26 & 4001224732336 & 337.217339 & -58.444885 & 0.955972 & No entry & Unreferenced \\
  27 & 4000781402752 & 216.968619 & 34.575819 & 0.974076 & No entry & Unreferenced \\
  28 & 4001283017901 & 120.202582 & 36.058927 & 0.994169 & [2016ApJS..224....1R] & Referenced \\
  29 & 4000833486119 & 116.260049 & 39.457642 & 0.971092 & No entry & Unreferenced \\
  30 & 4000949659908 & 146.342493 & 68.730869 & 0.961113 & No entry & Unreferenced \\
  31 & 4000982920478 & 53.084832 & -27.765379 & 0.983472 &  [2010A\&A...512A..12B] & Referenced \\
  32 & 4001189505548 & 192.492491 & 2.436292 & 0.992574 & No entry & Unreferenced \\
  33 & 4001060882070 & 89.700725 & -73.049783 & 0.962839 & No entry & Unreferenced \\
  34 & 4000889750512 & 151.176470 & 41.214096 & 0.962205 & No entry & Unreferenced \\
  35 & 6000322363510 & 53.149367 & -27.823945 & 0.963889 & [2016ApJ...830...51S] & Referenced \\
  36 & 4000722901091 & 28.257843 & -13.928090 & 0.982778 & No entry & Unreferenced \\
  37 & 6000198293960 & 264.488431 & 60.101798 & 0.986865 & No entry & Unreferenced \\
  38 & 4001095660911 & 258.587670 & 59.970358 & 0.955193 & No entry & Unreferenced \\
  39 & 4000972775076 & 330.960020 & 18.796346 & 0.989131 & No entry & Unreferenced \\
  40 & 4001132466571 & 126.545810 & 26.456196 & 0.997077 & No entry &  Unreferenced \\
  41 & 4000933395648 & 312.810365 & 2.288410 & 0.976252 & No entry & Unreferenced \\
  42 & 4000932940918 & 218.066960 & 32.997228 & 0.990737 & No entry & Unreferenced \\
  43 & 4001048433104 & 93.880689 & -57.754746 & 0.957755 & No entry & Unreferenced \\
  44 & 4001039919651 & 53.111470 & -27.673717 & 0.994424 & [2011ApJ...743..146C] & Referenced \\
  45 & 4001282607544 & 333.765783 & -14.006097 & 0.999520 & No entry & Unreferenced \\
  46 & 4000922341052 & 260.723839 & 58.849293 & 0.995477 & No entry & Unreferenced \\
  47 & 4000731518210 & 194.869144 & 14.146223 & 0.994651 & No entry & Unreferenced \\
  48 & 4001082523786 & 311.703084 & -12.869002 & 0.976454 & No entry & Unreferenced \\
  49 & 4000767041112 & 149.784518 & 2.172233 & 0.991335 & [2007ApJS..172...99C] & Referenced \\
  50 & 4001024667142 & 150.661685 & 1.718587 & 0.967865 & [2007ApJS..172...99C] & Referenced \\
  \enddata
\end{deluxetable*}

\begin{figure*}
  \includegraphics[width=\textwidth]{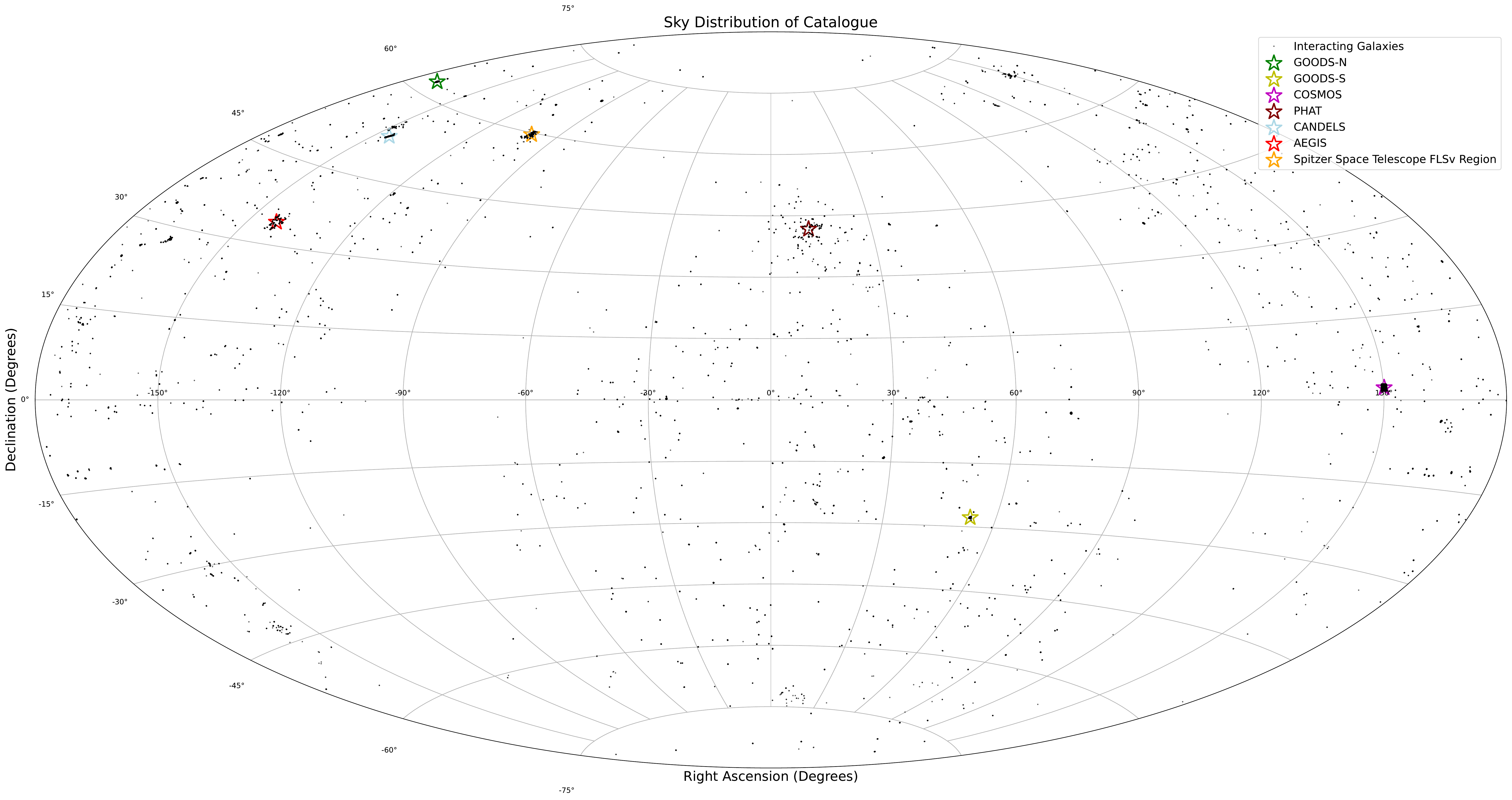}
  \caption{Sky Distribution of our catalogue, with marked positions of well known deep surveys conducted by \emph{HST}. \emph{HST} is able to observe almost the entire sky and therefore the interacting galaxies are scattered throughout. Large clusters of sources are found in the locations of surveys. This shows that often our sources are in the background of larger surveys and observations.}
  \label{fig:sky-distribution}
\end{figure*}

\vspace{-8mm}

\subsection{The Gems}\label{gems}
\noindent By conducting a visual inspection of the 27,720 candidate systems we were able to directly identify many other objects of astrophysical interest. As \texttt{Zoobot} was trained to highly predict objects with irregular morphologies, we also find many other astrophysical objects with strange morphologies which may be of interest to the community. We call these sources of contamination gems. We make 16 sub-categories of these: active galactic nuclei (AGN)/quasars, submillimetre galaxies, galaxy groups, high redshift galaxies, jellyfish galaxies, galaxy jets, gravitational lenses/lensing galaxies, Lyman-$\alpha$ Emitters, overlapping galaxies, edge on proto-planetary disks, radio halos, ringed galaxies, supernova remnants, transitional young stellar objects, young stellar clusters and unknown objects. 

Each sub-category has been defined by checking Simbad and ViZier for references within a 5$^{\prime \prime}$ radius of each source and using the astrophysical literature for a definition of the source. DOR classified any unreferenced objects by morphological similarity to other defined objects. The platforms ESASky\footnote{ESASky: \url{https://sky.esa.int/}}\citep{Merin_17}, NASA Extragalactic Database (NED) and the Sloan Digital Sky Survey were also used to investigate any unreferenced objects. ESASky was of paramount importance as we could investigate many objects across a range of wavelengths with many instruments. 

The only objects which were classified by other means than visual morphology were AGN/quasars, submillimetre galaxies and the six unknown objects. We attempt to confirm the unreferenced AGN/quasar as candidates by investigating the source in Chandra or XMM-Newton for hard or soft X-Ray emission. The submillimetre candidates were also investigated using Herschel or Planck measurements. If there was a positive signal in their positions, they were classified as such. Further work will be needed to confirm these classification.

The final category which required further inspection was that of the unknown objects. These are objects which have unusual morphology which mark them out from the rest of the sample, but no references associated with them in Simbad or ViZier. They also did not appear in NED, meaning they could not be confirmed to be galaxies. These objects are shown in appendix \ref{unknown-object}.

Table \ref{tab:gems} shows a breakdown of the total number of objects found and the number of which were referenced or unreferenced. We have released catalogues of each sub-category in the same format as that of the main catalogue without the interaction prediction column. Each of these catalogues can also be found at the same Zenodo link.

\begin{deluxetable*}{cccc}
  \tabletypesize{\footnotesize}
  \tablewidth{0pt}
  \tablecaption{A breakdown of gems found in the visual inspection stage of contamination. Each gem category has been classified based on the references associated with each object. \label{tab:gems}}
  \tablehead{
  \colhead{Category} & \colhead{Total Found} & \colhead{Referenced} & \colhead{Unreferenced}
  }
  \colnumbers
  \startdata
  AGN/Quasars & 35 & 21 & 14 \\
  Submillimetre Galaxies & 11 & 8 & 3 \\
  Galaxy Groups & 6 & 6 & 0 \\
  High Redshift Galaxies & 10 & 7 & 3 \\
  Jellyfish Galaxies & 18 & 5 & 13 \\\
  Galaxy Jets & 25 & 10 & 15 \\
  Gravitational Lenses/Lensing Galaxies & 189 & 64 & 125 \\
  Lyman-Alpha Emitters & 1 & 1 & 0 \\
  Overlapping Galaxies & 221 & 92 & 129 \\
  Edge-on Protoplanetary Disks & 9 & 2 & 7 \\
  Radio Halos & 1 & 1 & 0 \\
  Ringed Galaxies & 6 & 1 & 5 \\
  Supernova Remnants & 4 & 3 & 1 \\
  Transitional Young Stellar Objects & 2 & 1 & 1 \\
  Unknown Objects & 6 & 0 & 6 \\
  Young Stellar Clusters & 2 & 1 & 1 \\
  \enddata
\end{deluxetable*}

\vspace{-5mm}

\subsection{Source Redshifts and Photometry}\label{z_phot_analysis}
\noindent We investigate the redshift distribution and photometric properties of sources in our catalogue. We extract all sources with pre-existing data, querying Simbad, ViZieR, the HSC via the Milkulski Archive for Space Telescopes (MAST) and NED. Our queries use a $5^{\prime \prime}$ search radius within the Python package \texttt{AstroQuery}. The existing data from each of these databases has undergone heterogeneous selection and analysis procedures by the various studies we extract them from; we do not try to reconcile these here. Rather than a detailed physical analysis of these sources, our priority in this subsection is to highlight how to explore and use this catalogue, as well as any difficulties which may arise.

Of the 21,926 interacting systems in our high-confidence sample, 3,037 of the 7,522 referenced sources have a measured redshift. Figure \ref{fig:redshift-dist} shows the redshift distribution of this subset of our catalogue. 42.5\% of the sources have a redshift $z \leq 0.5$, 45.1\% have a redshift $0.5 < z < 1$ and 12.4\% have a redshift $z > 1$. In fact, a small fraction (15) of these sources are found to be at z$\geq 5$. Upon investigation of these sources two of their redshifts have been measured photometrically, while the remaining 13 sources did not have the method of measurement recorded in the archive. Therefore, this finding of very high redshift interacting galaxies are uncertain at best. 

It is important to note that the small sample with redshift information is affected by the selection biases of the combined studies publishing these values, and therefore the distribution may not be representative of the full sample. In addition, above redshift $z = 1$ the $F814W$ filter begins to only capture rest-frame UV flux, and therefore $z > 1$ galaxies with low star formation rates are more likely to fall below the flux limits of our detection images. Sampling only the rest-frame UV also changes a galaxy's observed brightness and morphology \citep[e.g.,][]{ferreira22} -- the latter being how \texttt{Zoobot} identifies interacting galaxies. For example, tidal features whose initial starburst has faded may be undetected; conversely, a single galaxy with irregular star-forming clumps may appear to be multiple interacting galaxies, which we noted as a particular source of contamination during the visual inspection stage. High-redshift interacting galaxies that are detected initially by \texttt{Zoobot} but have unusual morphologies compared to $z \sim 1$ sources may be removed during prediction (Section \ref{CNNs}), given that finetuning is based primarily on the $z \lesssim 1$ imagery of Galaxy Zoo: \emph{Hubble}. Therefore, the currently measured redshift distribution in Figure \ref{fig:redshift-dist} is likely due to some combination of selection bias and training bias. 

\begin{figure}
  \centering
  \includegraphics[width=\columnwidth]{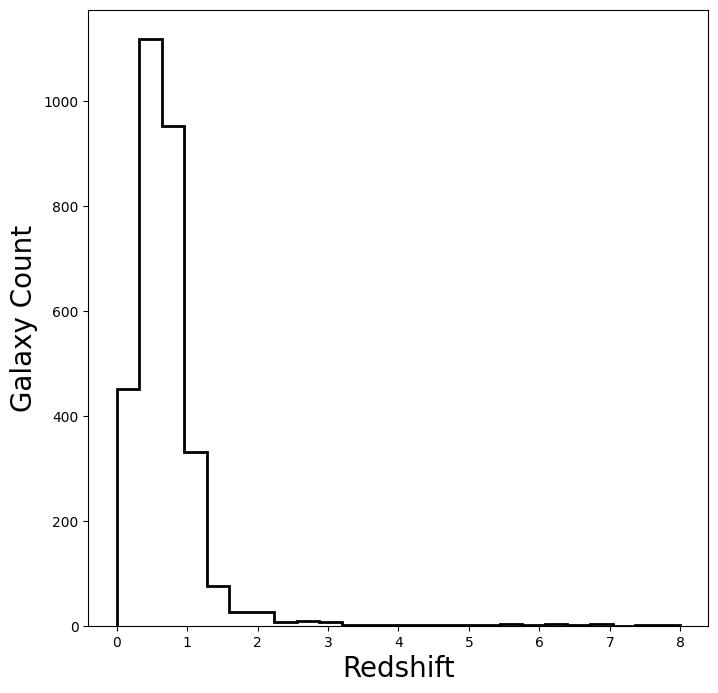}
  \caption{The redshift distribution of a subsample of our catalogue. Of the 7,583 referenced systems, 3,037 of them had redshift measurements in the NED, MAST or Simbad. This redshift distribution shows that our model confidently predicted interacting systems primarily for $z < 1$ systems. This was anticipated, as the model was primarily trained on systems at these redshifts. There are fifteen sources with a reported $z > 5$.}
  \label{fig:redshift-dist}
\end{figure}

\begin{figure}
    \centering
    \includegraphics[width=\columnwidth]{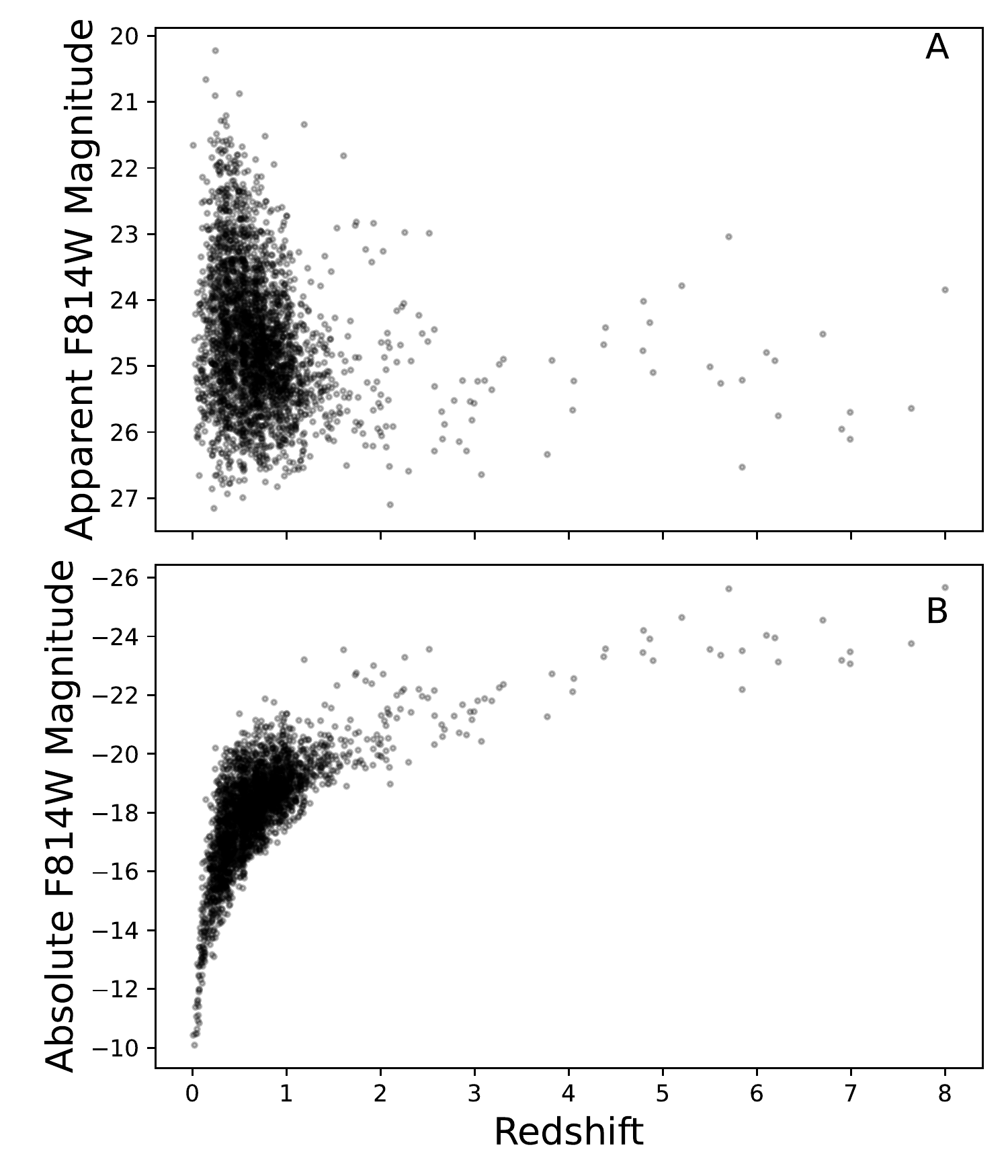}
    \caption{The distribution of redshift with magnitude for all sources with available data. This shows the parameter space we are sampling in this catalogue. Panel A shows that the majority of our sources are dim, background sources at low redshift. Panel B shows the faintest objects we find are at the limiting magnitudes of the different surveys this data is from.}
    \label{fig:redshift-mag-dist}
\end{figure}

Figure \ref{fig:redshift-mag-dist} shows the basic parameter space sampled by the sub-sample of the catalogue with existing photometry and redshifts. We show the distributions of redshift with the measured apparent $F814W$ magnitude and the calculated absolute $F814W$ magnitude. The faintest objects are, as expected, observed at approximately the limiting magnitude of the deepest observations in our catalog. Other observations have brighter limits; those wishing to select a uniform or volume-limited sample from our catalog must consider the variable flux limits across the sample.

We finally focus on sources from our high-confidence sample that have multi-band photometry, focusing on commonly-observed filters. By construction, 100\% of the sample has $F814W$ measurements, with 45\% of the catalogue having $F606W$ and only 11\% having measured fluxes in $F475W$. Table \ref{tab:filters-breakdown} summarizes the filter coverage of our catalogue. 6.1\% (1336 sources) have complete 3-band photometric information in the HSC. We use these to create examples of color images from the catalogue \citep[using the algorithm of][]{lupton_04}. We used a scaling factor $Q = 2$ and $\alpha = 0.75$, with ($F814W$, $F606W$, $F475W$) as RGB channels and multiplicative factors of (1.25, 0.95, 2). The resultant images are shown in Appendix \ref{colour-images}.

\begin{deluxetable}{cc}
  \tabletypesize{\small}
  \tablewidth{0pt}
  \tablecaption{Percent of sources in the final catalogue which have observations in the relevant \emph{Hubble} filter. \label{tab:filters-breakdown}}
  \tablehead{
  \colhead{Filter (s)} & \colhead{Sources Covered}
  }
  \colnumbers
  \startdata
  F814W & 100\% \\
  F606W + F814W & 45.0\% \\
  F475W + F814W & 11.0\% \\
  F475W + F606W + F814W & 6.1\% \\
  \enddata
\end{deluxetable}

We extract the measured magnitudes of the $F606W$ and $F814W$ filters, giving us two-band photometry for 9,876 sources. Cross referencing with each source that had a redshift yields 2,993 sources from our catalogue. We calculate the color of each source and plot it against the absolute magnitude in the $F814W$ filter. Figure \ref{fig:colour-magnitude} shows the resulting color-magnitude distribution in Panel A. The resultant distribution is very hard to interpret due to the high scatter of the sources. We extrapolate from this panel that there is little contamination from sources other than galaxies. If levels of contamination were high we would expect a second locus of sources with a very different color-magnitude distribution.

Plotting the color-magnitude distribution in this way captures a wide range of rest-frame wavelengths in the observed filters, which is the primary reason that panel A of Figure \ref{fig:colour-magnitude} is hard to interpret. In this first-look study, we do not have full spectral energy distributions (SEDs) of most sources, so K-correction of individual colors within this sample would involve assuming a template SED for each galaxy. Given that a high fraction of galaxies in our sample of mergers may deviate from standard SED templates, we wish to avoid this method. Instead, we choose redshift ranges within which to examine subsamples, such that the observed $F606W$ and $F814W$ bands cover consistent rest-frame colors within that subsample. Figure \ref{fig:colour-magnitude}B shows only sources with $z < 0.18$, within which the observed filters can be taken to be approximately rest-frame filters, which we define as at least 50\% of the flux captured in the observed band being emitted at rest-frame wavelengths covered by that band. At $0.24 < z < 0.56$, the observed $F606W$ filter captures at least 50\% rest-frame $F475W$ flux, and the observed $F814W$ filter captures at least 50\% rest-frame $F606W$ flux, so Figure \ref{fig:colour-magnitude}C is approximately a rest-frame $F475W - F606W$ vs $F606W$ plot. At $0.62 < z < 1$, Figure \ref{fig:colour-magnitude}D is approximately a rest-frame NUV-Blue plot ($F336W - F475W$ vs $F475W$). 

\begin{figure}
  \centering
  \includegraphics[width=\columnwidth]{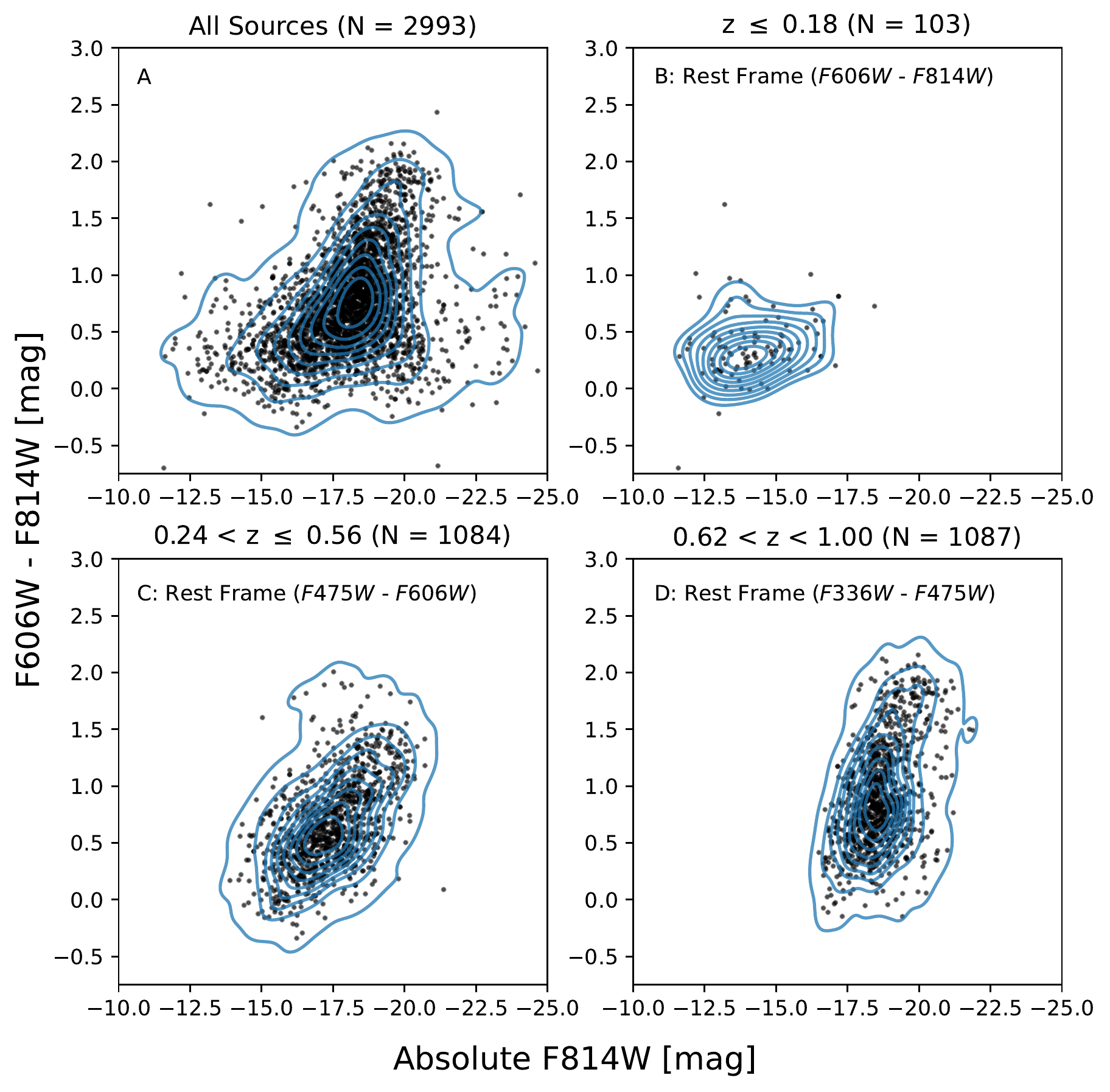}
  \caption{The color-magnitude distribution of sources with a redshift measurement associated. Panel A shows the distribution of all galaxies, without controlling for redshift or dust extinction. The remaining panels then split these sources into distinct redshift bins where the $F606W$ and $F814W$ filters are observing in different rest frames. Panel B shows the color-magnitude distribution in the local universe, where the rest frame observations are $F606W$ and $F814W$ flux. This bin reveals a blue population. Panel C shows the redshift bin where at 50\% - 100\% of observed $F606W$ and $F814W$ flux is rest frame $F475W$ and $F606W$ flux. This bin reveals a larger distribution of interacting galaxies, with a dominating population of blue systems and a minor population of red systems. Panel D shows the redshift bin where 50\% to 100\% of observed $F606W$ and $F814W$ flux is rest frame $F336W$ and $F475W$ flux. These filter bands are very sensitive to star formation, and reveal a broad distribution in color of red and blue systems.}
  \label{fig:colour-magnitude}
\end{figure}

The galaxies in Panel B are observed in approximately the rest frame $F606W$ and $F814W$ filters. Nearly all are blue systems \citep[by general definitions at various redshifts, \emph{e.g.},][]{kauffmann03, whitaker12, schawinski14}. This is expected for interacting systems with enough gas to fuel a starburst. The lack of many red systems is due to few gas-poor (``dry'') interactions in the (relatively) local volume \citep{lopez09}. In Figure \ref{fig:colour-magnitude}C, the $F606W$ and $F814W$ filters are still detecting rest-frame optical ($F475W$ and $F606W$) emission, and we find a much broader population. There are both blue and red interacting systems, with the redder mergers occurring in more luminous (likely higher mass) systems, broadly consistent with expectations \citep{vandokkum05, lotz08}. The rest-frame filters approximately captured in Panel D ($F336W$ and $F475W$) sample emission across the $4000$\,\AA\ break. Sensitivity to NUV means this panel effectively splits systems according to very recent star formation history \citep{schawinski14,smethurst15}. There is a significant spread in color, with equivalent red and blue systems. We, therefore, find many young blue systems undergoing star formation and bright brighter, elliptical, massive systems also undergoing interaction in this bin.

This initial examination of the subsample of systems with easily retrievable redshifts has revealed that the interacting galaxies in the sample broadly agree with previous studies of colors in merging systems. This demonstrates the underlying promise of the catalogue. A detailed study is beyond the scope of this work, but there is considerable potential for new astrophysical insights using this high-confidence catalog with nearly an order of magnitude more sources than those previously published.

\vspace{-3mm}

\section{CONCLUSION}\label{conclusion}
\noindent We present a large, pure catalogue of 21,926 interacting galaxy systems found from the \emph{Hubble} Source Catalogue. This catalogue is a factor of six larger than previous works. Each interacting system was found using the European Space Agency's new platform ESA Datalabs, which allowed us to directly apply an advanced CNN - \texttt{Zoobot} - to the entire \emph{Hubble} science archive. This corresponds to predicting over 126 million sources. The compiled catalogue has a contamination rate of $\approx$3\% as found by bootstrapping. Table \ref{tab:ex-cat} shows an example of 50 entries in our new catalogue, Figure \ref{fig:interactors} showing the corresponding images. The new catalogue and all corresponding images can be downloaded from Zenodo: \emph{Zenodo link will be added}.

Each of our interacting galaxies were given a prediction score $\geq$0.95 by \texttt{Zoobot}, with such a conservative score chosen to limit contamination and maintain purity in the catalogue. Contamination was removed by applying cuts in representation space (shown by Figure \ref{fig:cuts-visual}) and visual inspection. Upon visual inspection, many contaminating images were found to be objects of other astrophysical interest. These have been compiled into separate catalogues, and Table \ref{tab:gems} shows a breakdown of the objects found. These sub-catalogues have been released alongside our interacting galaxy catalogue. With the priority of purity in this catalogue creation, we will aim in future work to use it in the statistical analysis of interacting galaxies and begin linking the underlying parameters of interaction to the complex physical processes that occur in them. A secondary purpose of this catalogue is to serve as a training set for future models which may wish to search for interacting or merging galaxies.

With the use of ESA Datalabs, this project was conducted quickly. The entire process, from creating the source cutouts, to training \texttt{Zoobot}, to making predictions on 126 million sources took three months to complete. Using conventional methods, such as \texttt{AstroQuery} or TAP services, downloading the data would have likely taken on this timescale. By bringing the user to the data, rather than vice versa, catalogues of a similar size - and many times larger than previous catalogues - of many different objects can be created quickly.

None of the the interacting systems in this work are `new'; every one of them exists in the background of large scale \emph{HST} surveys and observations since their release. However, the method to directly search for them has been impractical until the release ESA Datalabs. By directly applying machine learning to existing astrophysical data repositories, a new method to creating significantly larger catalogues has been achieved.

This shows the importance of archival work, and the power that ESA Datalabs will bring to the field of astronomy. ESA Datalabs is expected to be released in Q3 and with it, the ability for large scale exploration of archival data. It will be released with introductory tutorials, step-by-step guides and different Python environments for ease of use for different telescopes and instruments the ESA is involved in. It will have a full cluster of GPUs at its disposal and a storage capability in the range of hundreds of Terabytes. In future, this entire project - from training set creation to predictions - could be conducted on ESA Datalabs.

Such a setup as ESA Datalabs also allows the creation of large observational catalogues, comparable to that we create from cosmological simulations. This is incredibly important to further constraining already existing results. In the current period of astronomy where large survey instruments are awaiting first light, or the beginning of future telescopes is uncertain, the ability to get ever more information out of the archives is paramount.

\section*{ACKNOWLEDGEMENTS}
\noindent DOR gratefully acknowledges the support from European Space Agencies Visitor Archival Research program, and hosting at the European Space Astronomy Centre. DOR thanks Bruno Mer\'in for supervising this project and Sarah Kendrew for aiding its creation. This project was conducted as part of DORs PhD program supported by the UK Science and Technology Facilities Council (STFC) under grant reference ST/T506205/1. BDS acknowledges support through a UK Research and Innovation Future Leaders Fellowship [grant number MR/T044136/1]. ILG acknowledges support from an STFC PhD studentship [grant number ST/T506205/1] and from the Faculty of Science and Technology at Lancaster University. MW gratefully acknowledges support from the UK Alan Turing Institute under grant reference EP/V030302/1. MRT acknowledges the support from an STFC PhD studentship [grant number ST/V506795/1] and from the Faculty of Science and Technology at Lancaster University.

Much of the intense computation was conducted at the High End Computing facility at Lancaster University. This publication uses data generated via the Zooniverse.org platform, and the unending enthusiasm of citizen scientists and volunteers in classifying galaxies. We also thank the many PIs who's archival data we have used to create this catalogue. All data containing astrophysical objects of interest found in this work are public on MAST: \dataset[10.17909/wfke-n133]{http://dx.doi.org/10.17909/wfke-n133}.

This research made use of many open-source Python packages and scientific computing systems. These included \texttt{Matplotlib} \cite{matplotlib}, \texttt{scikit-learn} \citep{scikitlearn}, \texttt{scikit-image} \citep{scikitimage}, \texttt{Pandas} \citep{pandas}, \texttt{Shapely} \citep{shapely}, \texttt{UMAP} \citep{umap} and \texttt{numpy} \citep{numpy}. This work also extensively used the community-driven Python package Astropy \citep{astropy:2018}. \texttt{Zoobot} utilises the underlying code \texttt{Tensorflow} \citep{tensorflow} Python package.

This project used data from the \emph{Hubble} Space Telescope and stored in the archives at the European Space Astronomy Centre. These observations are obtained from the Space Telescope Science Institute, which is operated by the Association of Universities for Research in Astronomy, Inc, under NASA contract NAS 5-26555. All sources were found using v3.1 of the \emph{Hubble} source catalogue \citep{whitmore_16} and accessed using the ESA Datalabs science platform. ESA Datalabs is directly connected to the ESA \emph{Hubble} Science Archive. This study makes use of data from AEGIS, a multiwavelength sky survey conducted with the Chandra, GALEX, Hubble, Keck, CFHT, MMT, Subaru, Palomar, Spitzer, VLA, and other telescopes and supported in part by the NSF, NASA, and the STFC.

 For the purpose of open access, the authors have applied a Creative Commons Attribution (CC BY) licence to any Author Accepted Manuscript version arising.

DOR would like to thank those in the ESA Traineeship program cohort of 2022. They created a wholly welcoming environment and space of support. A special thanks must go to Karolin Frohnapfel and Emma Vellard for much technical discussion. Finally, DOR would like to acknowledge Aur\'elien Verdier.

\bibliographystyle{aasjournal}

\appendix

\vspace{-8mm}

\section{Further Model Diagnostics}\label{further-diagnostics}
\noindent In Section \ref{diagnostics} we present diagnostic properties of our model. These include the accuracy measurements, purity measurements as well as confusion matrices at different cutoffs of our model. Here, we present the Receiver Operating Characteristic (ROC) curves, the precision-recall (PR) curves, and measures of true and false positive rates vs the cutoff threshold.

Figure \ref{fig:pr-roc-curves} shows the ROC and PR curves of the final \texttt{Zoobot} model we applied to the the \emph{Hubble} archives. The ROC shows the rate of change of finding true positives and false positives with changing cutoff. The PR curve shows the changes of precision against recall. Precision is the ratio of true positives (interacting galaxies correctly predicted as so) to the sum of true and false positives (non-interacting galaxies incorrectly predicted as interacting). The recall is then the ratio of true positives to the sum of true positives and false negatives (interacting galaxies that have been misclassified as non-interacting). The red crosses in both plots shows how the model was behaving when we use a cutoff of 0.95. 

These are both as expected. Both curves show that the model behaves well, and are much better than a random classifier (which would have a 1:1 relation). The ROC plot shows that we are minimising our false positive rate when using a prediction score cutoff of 0.95. However, we are misclassifying approximately 50\% of interacting galaxies as non-interacting galaxies. The contamination rate in our final catalogue (False Positives rate) will be very low (close to zero in this ideal validation set). The PR curve shows a similar result. Here, we are operating with a high precision (finding a pure catalogue) while keeping our recall minimal. 

We also present the changing F1 score for the model used in this work, shown in Figure \ref{fig:f1-score}. The F1 score is twice the ratio of precision multiplied by recall upon precision summed to recall. This combines our measure of accuracy and purity into a single metric. The cutoff we use in this work is at the point where the F1 score has began to decline. This is because we are beginning to lose recall rapidly, but gaining significantly in precision. As discussed in Section \ref{diagnostics}, this was an acceptable trade off in this work for a very large, pure interacting galaxy catalogue.

\begin{figure*}
  \centering
  \includegraphics[width = 0.95\textwidth]{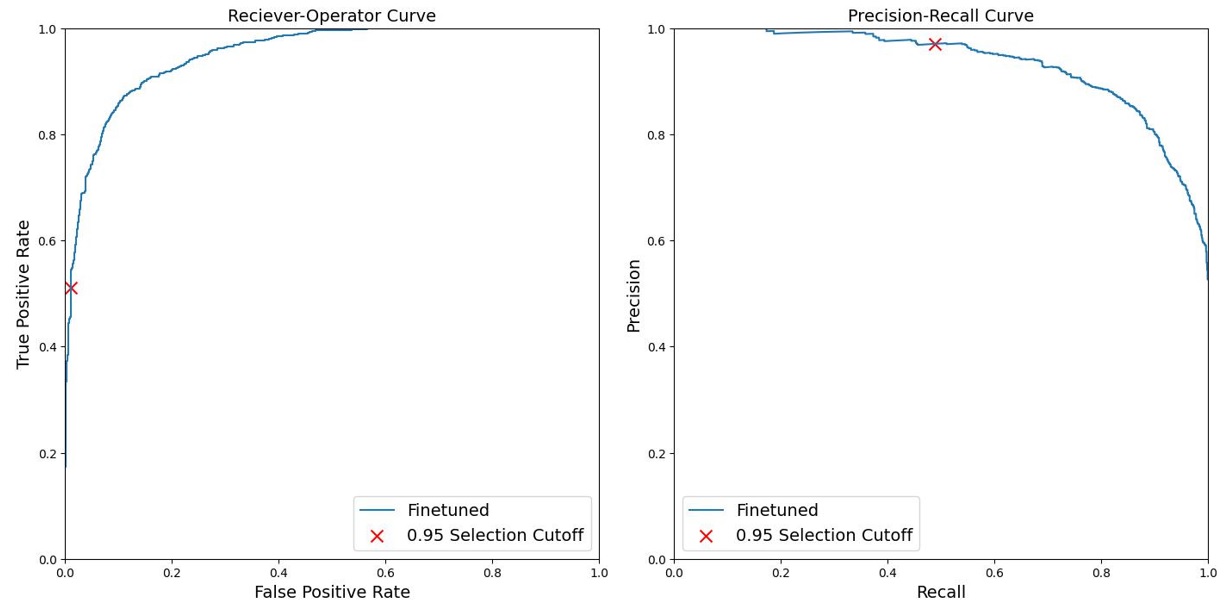}
  \caption{The Receiver-Operator and Precision-Recall Curve for the \texttt{Zoobot} model that was used to explore the Hubble archives. The blue curves are the measured curves. These curves measure the relevant rates or characteristics based on the changing cutoff applied to how \texttt{Zoobot} defines an interacting galaxy. The red crosses are where the prediction score cutoff is for this work. We can see in the Reciever-Operator Curve that the prediction score cutoff we use would have an incredibly low false positive rate, while it would be misclassifying $\approx$50\% of interacting galaxies. This also shown in the precision recall curve where our recall is $\approx$50\%.}
  \label{fig:pr-roc-curves}
\end{figure*}

\begin{figure*}
    \centering
    \includegraphics[width = 0.65\textwidth]{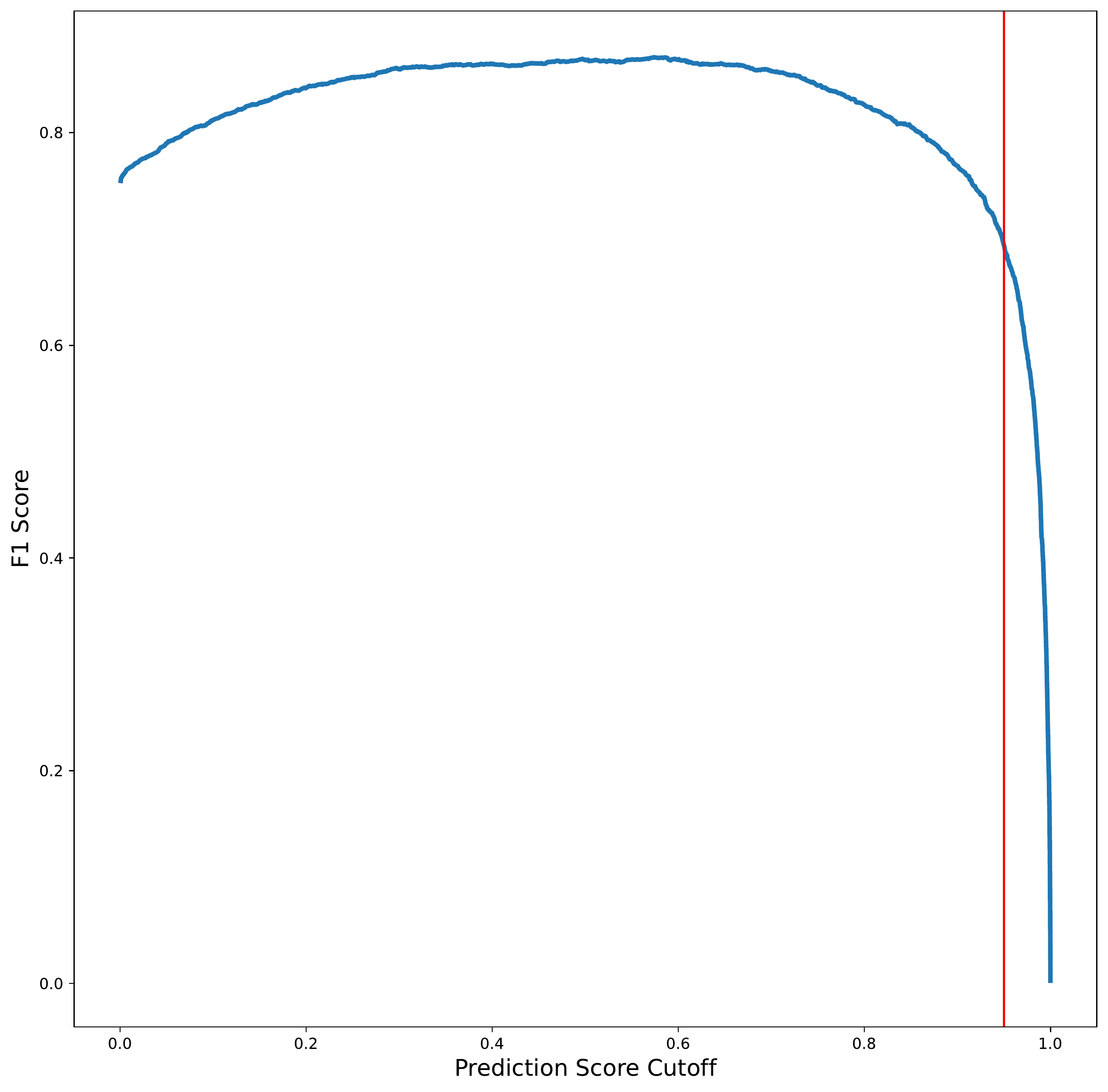}
    \caption{The F1 score found during the diagnostics of the model used in this work. The F1 score is a measure combing the measure of accuracy and purity into one metric. The cutoff we use is at the point where the F1 score begins to rapidly decline. This point is shown by the red vertical line.}
    \label{fig:f1-score}
\end{figure*}

\section{Examples of Sources with 3-Band Information}\label{colour-images}
\noindent Of the full catalogue of 21,926 interacting systems, only 1336 of them had got all 3-band information. Six examples are shown in Figure \ref{fig:colour-images}. These were created using the \citet{lupton_04} algorithm, with a scaling factor $Q = 2$ and $\alpha = 0.75$, with ($F814W$, $F606W$, $F475W$) as RGB channels and multiplicative factors of (1.25, 0.95, 2).

\begin{figure*}
  \centering
  \includegraphics[width = 0.85\textwidth]{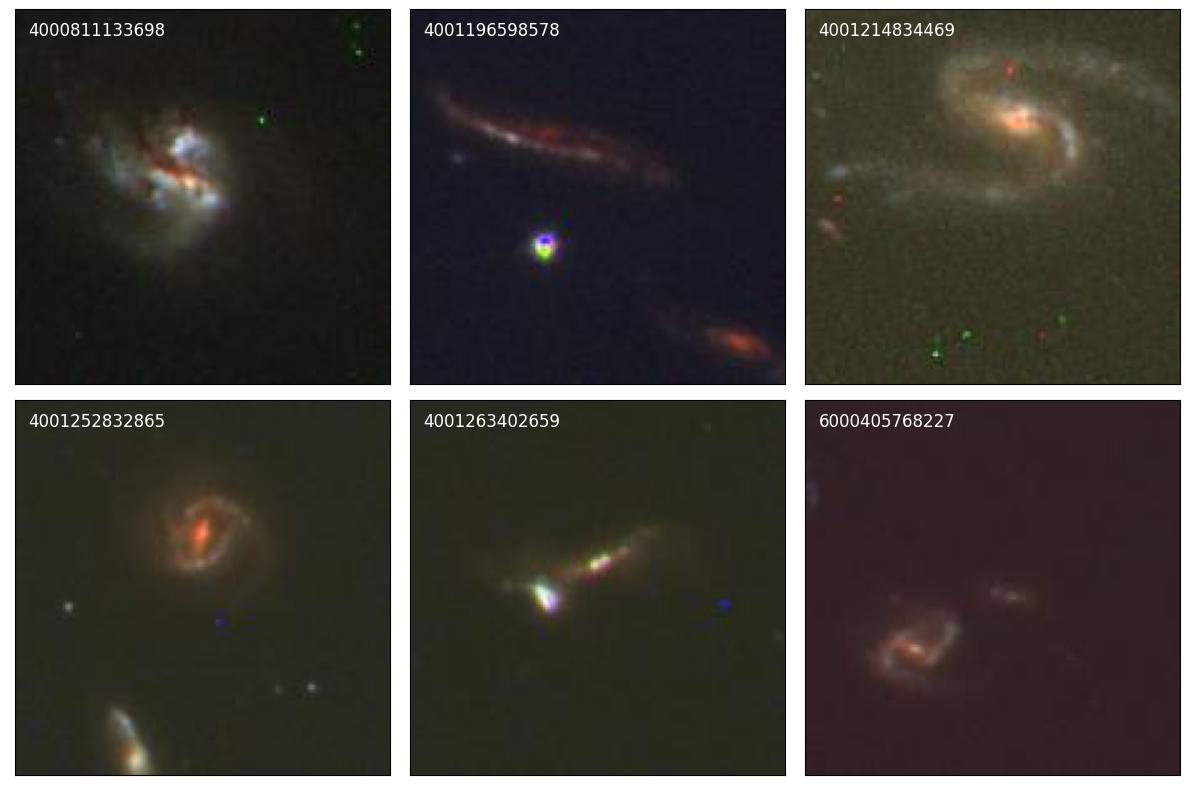}
  \caption{Example of six interacting systems in the catalogue with full 3-band imagery.}
  \label{fig:colour-images}
\end{figure*}

\section{Unknown Objects}\label{unknown-object}
\noindent From the final catalogue, there were six sources which we could not visually identify. These objects were also not referenced anywhere in the astrophysical literature. $F814W$ cutouts of the six objects are shown in Figure \ref{fig:unknown-images}. Their Source IDs are shown in the upper left of each image, and a separate catalogue has been released of these with all other objects. This catalogue can be found at the data release on Zenodo.

Four of the six objects (40001156424176, 4001368788120, 4001418076626 and 6000398415347) have a bright central source, followed by a low-surface brightness tail. Initially, it was assumed that these were solar system objects such as comets. This, however, could not be confirmed. The first of these four sources is also thought to potentially be a highly disruped system with a significantly elongated tidal feature. The final two unknown sources (6000186797547 and 6000341449179) have no clear central source, though there is extended structure to them. These are likely to be highly irregular galaxies, but no confirmation could be found.

These objects are released to the community for identification and investigation, as the authors cannot find definitive agreement on what they are. 

\begin{figure*}
  \centering
  \includegraphics[width = 0.85\textwidth]{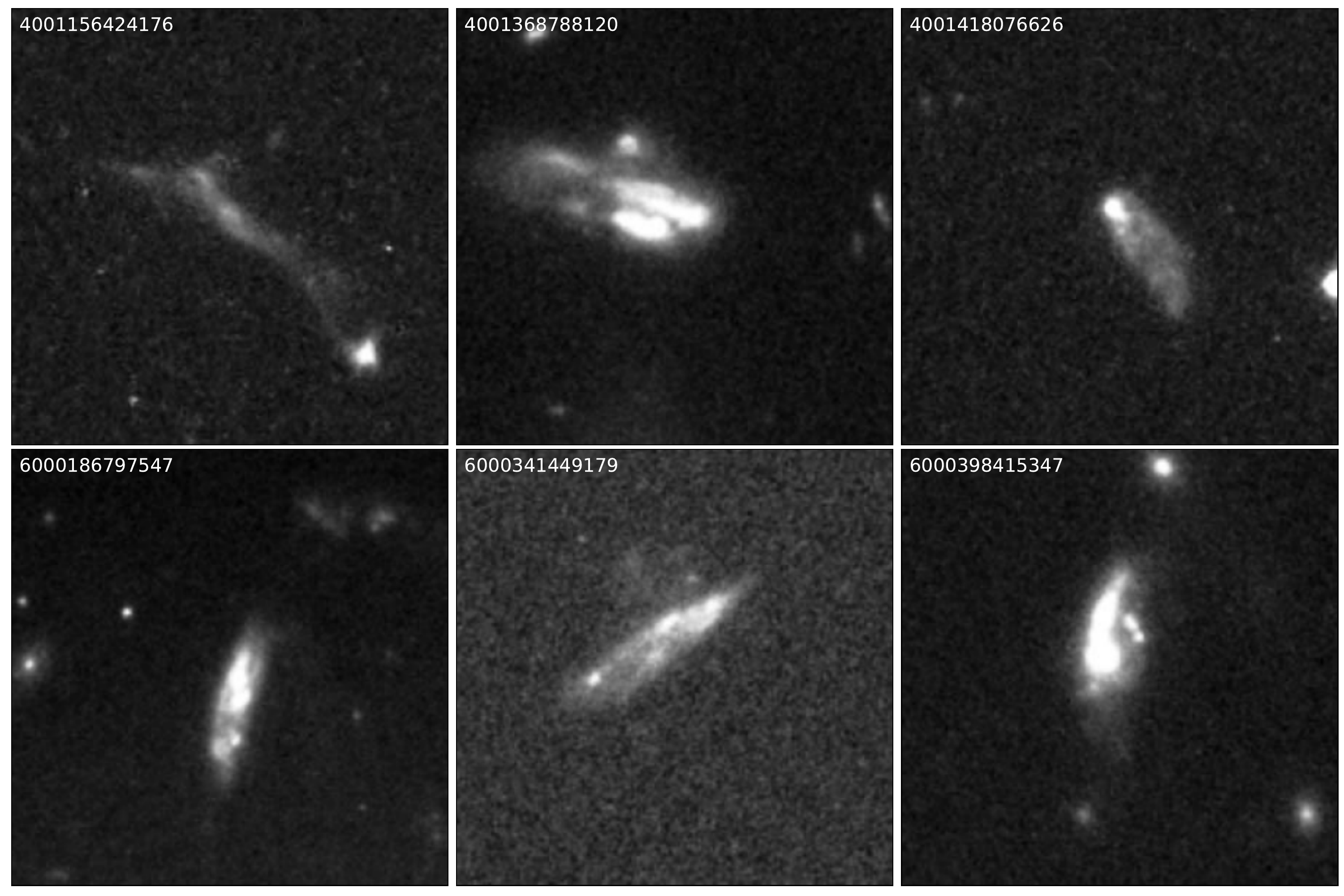}
  \caption{The six unknown systems found in this work. These have no reference in Simbad or in NED, and their morphology could not be classified by the authors. Investigation into these six objects are presented to the community, with the authors hoping that future work and investigation of them can be conducted by them.}
  \label{fig:unknown-images}
\end{figure*}

\section{Acknowledging PIs}
\noindent In the final section of this work, we wish to acknowledge all of the PIs whose observations we have used. A machine readable table containing the proposal IDs, the DOIs and the references (if provided/found) is presented with this work. Table \ref{tab:pis} shows the first twenty observations used in this work and is an example of this table.

\begin{deluxetable*}{ccccc}
\tabletypesize{\small}
\tablewidth{0pt}
\tablecaption{Twenty example of the accompanying data table of observations used. \label{tab:pis}}
\tablehead{
\colhead{Proposal ID} & \colhead{Observation ID} & \colhead{Observation Date} &        \colhead{DOI} & \colhead{references}
}
\colnumbers
\startdata
    8183       &   hst\_8183\_54\_acs\_wfc\_f814w\_j59l54 &       18/07/2002 &    https://doi.org/10.5270/esa-88k8vcj & \\
    9075       &   hst\_9075\_2a\_acs\_wfc\_f814w\_j6fl2a &       24/07/2002 &    https://doi.org/10.5270/esa-gsxhb4b & \\
    9351       &   hst\_9351\_11\_acs\_wfc\_f814w\_j8d211 &       31/03/2003 &    https://doi.org/10.5270/esa-5lba8bo & \\
    9361       &   hst\_9361\_03\_acs\_wfc\_f814w\_j8d503 &       22/07/2003 &    https://doi.org/10.5270/esa-ecmnqgh & \\
    9363       &   hst\_9363\_09\_acs\_wfc\_f814w\_j8d809 &       02/07/2002 &    https://doi.org/10.5270/esa-ethtec5 & \\
    9367       &   hst\_9367\_02\_acs\_wfc\_f814w\_j8ds02 &       10/06/2003 &    https://doi.org/10.5270/esa-3j404ll & \\
    9373       &   hst\_9373\_02\_acs\_wfc\_f814w\_j6la02 &       05/07/2002 &    https://doi.org/10.5270/esa-ztsq94u & \citet{Rejkuba_05}\\
    9376       &   hst\_9376\_02\_acs\_wfc\_f814w\_j8e302 &       13/07/2002 &    https://doi.org/10.5270/esa-h90iavd & \citet{Keel06}\\
    9381       &   hst\_9381\_02\_acs\_wfc\_f814w\_j8fu02 &       13/03/2003 &    https://doi.org/10.5270/esa-vlapyea & \\
    9400       &   hst\_9400\_04\_acs\_wfc\_f814w\_j6kx04 &       29/05/2003 &    https://doi.org/10.5270/esa-39rnout & \\
    9403       &   hst\_9403\_02\_acs\_wfc\_f814w\_j8fp02 &       09/07/2002 &    https://doi.org/10.5270/esa-k5mv9ct & \\
    9405       &   hst\_9405\_6k\_acs\_wfc\_f814w\_j8iy6k &       22/05/2003 &    https://doi.org/10.5270/esa-zy9phm1 & \\
    9409       &   hst\_9409\_03\_acs\_wfc\_f814w\_j6n203 &       29/06/2003 &    https://doi.org/10.5270/esa-vjngw7r & \citet{goudfrooij04}\\
    9411       &   hst\_9411\_09\_acs\_wfc\_f814w\_j8dl09 &       11/02/2003 &    https://doi.org/10.5270/esa-debpiln &  \\
    9427       &   hst\_9427\_13\_acs\_wfc\_f814w\_j6m613 &       21/10/2002 &    https://doi.org/10.5270/esa-bw1b97v &  \\
    9438       &   hst\_9438\_01\_acs\_wfc\_f814w\_j6me01 &       16/01/2003 &    https://doi.org/10.5270/esa-e5eaam5 & \citet{gregg17}\\
    9450       &   hst\_9450\_02\_acs\_wfc\_f814w\_j8d402 &       25/08/2002 &    https://doi.org/10.5270/esa-9ttmykz & \citet{york05}\\
    9453       &   hst\_9453\_02\_acs\_wfc\_f814w\_j8f802 &       03/12/2002 &    https://doi.org/10.5270/esa-1xvyjfy & \citet{brown03}\\
    9454       &   hst\_9454\_11\_acs\_wfc\_f814w\_j8ff11 &       23/03/2003 &    https://doi.org/10.5270/esa-xsdowj9 &  \\
\enddata
\end{deluxetable*}

\label{lastpage}

\end{document}